\documentclass[pra,aps,groupedaddress,floatfix,twocolumn,superscriptaddress,showpacs,nofootinbib]{revtex4-1}
\usepackage{amsmath,amssymb,epsfig,bm,pifont}
\usepackage[linkcolor=black,citecolor=black,urlcolor=blue,colorlinks=true,linktocpage=true]{hyperref}
\usepackage{graphicx}
\usepackage{epstopdf}
\usepackage{bm}
\usepackage{tabularx}
\usepackage[utf8]{inputenc}
\usepackage[T1]{fontenc}
\usepackage{color}
\newcommand{\nn}{\nonumber }

\newcommand{\be}{\begin{eqnarray}}
\newcommand{\ee}{\end{eqnarray}}

\newcommand{\beq}{\begin{equation}}
\newcommand{\eeq}{\end{equation}}
\newcommand{\bea}{\begin{eqnarray}}
\newcommand{\eea}{\end{eqnarray}}
\newcommand{\tr}{\mathrm{tr}}

\newcommand{\veps}{\varepsilon}

\newcommand{\kf}{k_\textnormal{F}}

\begin{document}

\title{{Evolution from few- to many-body physics in one-dimensional Fermi systems:} \\
One- and two-body density matrices and particle-partition entanglement}

\author{Lukas Rammelm\"uller}
\email{lrammelmueller@theorie.ikp.physik.tu-darmstadt.de}
\affiliation{Department of Physics and Astronomy, University of North Carolina, Chapel Hill, North Carolina, 27599, USA}
\affiliation{Institut f\"ur Kernphysik (Theoriezentrum), Technische Universit\"at Darmstadt,
D-64289 Darmstadt, Germany}

\author{William J. Porter}
\email{wjporter@email.unc.edu}
\affiliation{Department of Physics and Astronomy, University of North Carolina, Chapel Hill, North Carolina, 27599, USA}

\author{Jens Braun}
\email{jens.braun@physik.tu-darmstadt.de}
\affiliation{Institut f\"ur Kernphysik (Theoriezentrum), Technische Universit\"at Darmstadt,
D-64289 Darmstadt, Germany}
\affiliation{ExtreMe Matter Institute EMMI, GSI, Planckstra{\ss}e 1, D-64291 Darmstadt, Germany}

\author{Joaqu\'{i}n E. Drut}
\email{drut@email.unc.edu}
\affiliation{Department of Physics and Astronomy, University of North Carolina, Chapel Hill, North Carolina, 27599, USA}

\begin{abstract}
{We study the evolution from few- to many-body physics of fermionic systems in one spatial dimension
with attractive pairwise interactions. We} determine the detailed form of the
momentum distribution, the structure of the one-body density matrix, and the pairing properties encoded in the
two-body density matrix. {From the low- and high-momentum scaling behavior of the single-particle momentum distribution we estimate
the speed of sound and {\it Tan}'s contact, respectively. Both quantities are found to be in agreement with previous calculations.
Based} on our calculations of the one-body
density matrices, we also present results for the particle-partition entanglement entropy, for which {we find a logarithmic
dependence on the} total particle {number.}
\end{abstract}

%\date{\today}
\maketitle

%%%%%%%%%%%%%%%%%%%%%%%%%%%%%%%%%%%%%%%%%%%%%
\section{Introduction}

In the last decade, the detailed experimental study of ultracold atoms has revealed a
stunning array of phenomena {in a wide range} of situations.
The ever-growing ability to control and measure the properties of atomic clouds has enabled the study of fermionic and bosonic systems in
one, two, and three dimensions, lattices, and even mixed dimensions
{by carefully tuning external trapping potentials (see, e.g., Refs.~\cite{coldAtomsReview, expTrap1D2D, Nature1DTraps}).}
As is well known, that versatility extends to changes in the interaction strength via
{\it Feshbach} resonances~\cite{JETFeshbach} as well
as to control over the degree of polarization (pseudo-spin population imbalance) and mass asymmetry (by isotopic variations).
Such an unprecedented plasticity makes the investigation of the challenging strongly coupled regimes both urgent {and necessary.}

On the computational side, the quantitative
characterization of these quantum few- and many-body systems poses a formidable challenge. The complexity of quantum many-body
physics presents itself in two ways. In non-stochastic methods, such as exact diagonalization, the memory requirements scale
{factorially} with the size
of the system (number of particles or spatial volume, depending on the algorithm), simply
because that is how the dimension of the
{\it Hilbert} space grows.
{In stochastic methods, namely quantum Monte Carlo (QMC) and all its incarnations, full access to
eigenstates is relaxed in exchange for answers to specific
questions (i.e. specific
correlation functions) and so the memory limitations are much milder.

As far as computations of properties of 1D systems are concerned, there}
are several non-stochastic ways to arrive at the ground-state properties, such as the {\it Bethe} ansatz~\cite{2013RvMP...85.1633G},
(density-matrix) renormalization group approaches~\cite{Voit,*RevModPhys.77.259}, exact diagonalization~\cite{PhysRevA.88.033607,*0295-5075-109-2-26005,*PhysRevA.94.042118}, methods based on effective interactions~\cite{1367-2630-16-6-063003}
as well as coupled-cluster approaches frequently applied in quantum chemistry~\cite{1367-2630-17-11-115001}.

{One of the main objectives of the present work is to set
benchmarks that show quantitatively and in detail what is presently feasible with our lattice QMC
approach. To this end, we
expand on previous works~\cite{GCS1D,Loheac:2015fxa} and consider a model for
two-component {\it Fermi} gases in one dimension with an attractive contact interaction between the two species
(see Sec.~\ref{sec:model} for a brief discussion), focusing on
%However, unlike Ref.~\cite{Loheac:2015fxa}, where the equation of state of 1D spin-imbalanced {\it Fermi} gases
%was studied with the same MC approach, we} consider
systems with an equal population of
spin-up and spin-down fermions.}
The algorithm we use relies on an auxiliary field transformation to decouple
the spin species in the density channel. This allows for a relatively simple calculation of the
one-body density matrix as well as the on-site two-body density matrix (i.e. the pair-correlation
function). However, the latter
requires taking the square of the one-body density matrix for each auxiliary field configuration,
which increases the statistical noise significantly. {Therefore,
better statistics are needed to obtain a more accurate estimate for this quantity, which can be most easily tested and achieved in the one-dimensional (1D) limit.}

{In Sec.~\ref{sec:res}, we present fully non-perturbative calculations of the
one- and two-body density matrices along with the associated momentum distributions for our 1D system of
fermions.
We compute these
quantities in the ground state for different particle content and across a number of coupling strengths ranging from the
noninteracting case to the strongly-coupled regime.
The fact that this specific problem is in principle exactly solvable by way of
the {\it Bethe} ansatz~\cite{betheAnsatz} is of great importance to us as
it allows to benchmark our lattice QMC approach, e.g. with respect to the computation of
the ground-state energy and the speed of sound.
Since the exact analytic approaches successfully employed over decades in the 1D case
do not possess a straightforward generalization to higher-dimensional systems,
our present study aims to set the methodological stage for future
quantitative studies of one- and two-body density matrices
of {\it Fermi} gases in higher dimensions. {In particular, for future studies of systems with a finite spin- and mass-imbalance,
the computation of general correlation functions is of interest as the formation of an
inhomogeneous ground state is expected to leave its imprint on these quantities, see, e.g., Refs.~\cite{PhysRevA.78.033607,Lee:2011zzo,Roscher:2016chy}.
Although our present 1D study does not fully generalize to higher dimensions, e.g. with respect to the scaling of the observables with the lattice size, we still consider it useful from a methodological point of view as it is possible to take vastly more data than in higher dimensions. This allows us to test how our algorithms perform
against the background of the existing exact results in the 1D limit.

From a phenomenological point of view, 1D {\it Fermi} gases allow to study the transition from few- to many-body physics in detail, both
experimentally~\cite{2013Sci...342..457W,*2013PhRvL.111q5302Z} and theoretically~(see Ref.~\cite{2016EPJWC.11301002Z} for a review).
The exploration of this transition has been enriched by
the concepts of quantum information, such as entanglement, entanglement entropy, mutual information,
as well as corresponding methods to determine them.
We rush to add that a direct comparison of our present results to experimental data would be qualitative at best. Indeed,
we shall consider fermions in a box with periodic boundary conditions {since they have been found
to minimize finite-volume effects in relativistic model studies in the sense that the infinite-volume limit is approached faster than
in case of antiperiodic boundary conditions (see, e.g., Ref.~\cite{Braun:2005gy}). Of course, the latter are only one
representative of boundary conditions which do not allow for a zero-momentum mode. A
rigorous proof of this observation for general boundary conditions
and observables is
difficult, if possible at all.
In any case, for a quantitative comparison with experimental data in the future (in particular with respect to the few-body limit),
the} numerical implementation of trap potentials~\cite{Berger2014, Berger2015} as used in experiments will be required.
For example, harmonic traps are often
considered in 1D experiments~\cite{2013Sci...342..457W,*2013PhRvL.111q5302Z}. Interestingly, we note that
the realization of flat-bottom traps in experiments has now also become possible (see e.g.~\cite{Hueck2017}).
Compared to systems with periodic boundary conditions, however,
the computational cost at fixed system size for studies involving such trapping geometries has been found
to increase~\cite{Berger2015,McKenney:2015gba} in case of our present MC approach.}

{
As a direct and nontrivial application of our results for the density matrix,
we are able to push further and explore specific quantum-information aspects of this system by
computing the one-particle partition-entanglement entropy.
The so-called {\it R\'enyi} entanglement entropy has been a center of attention for the last few years as a possible order parameter for topological phase
transitions~\cite{Amico:2007ag,*RevModPhys.81.865}.
Indeed, it was found that the so-called area-law violation (specifically, a logarithmic modification to the expected
area law scaling with sub-system size) could signal such a phase change~\cite{AreaLaws}.
Our motivation for considering the particle partition form of the {\it R\'enyi} entanglement
entropy is based on the recent interest in this quantity
and its scaling with the system size, which has been empirically found to follow a logarithmic law as in the case of spatial
entanglement~\cite{1751-8121-42-50-504012,*PhysRevB.76.125310,*PhysRevLett.98.060401, PhysRevA.78.042326, 2017arXiv170310587B}.
While conventional studies of entanglement analyze the degree of spatial entanglement of a
sub-system, circumscribed by a specific region of space, with the rest of the system, the kind of
entanglement we study here is different. Particle-partition entanglement quantifies the degree of quantum
correlation between a sub-set of particles (identified by labels of the density matrix) and the rest of the particles
in the system. As in the case of spatial entanglement, the quantum correlation being measured includes statistical effects,
which are non-trivial even for non-interacting systems. However, particle-partition entanglement features
non-universal coefficients in the leading logarithms that vary with the particle statistics as well as with
the interaction strength, which makes it a useful diagnostic tool for quantum correlations in strongly coupled
matter~\cite{PhysRevA.78.042326, 2017arXiv170310587B}.
}

%%%%%%%%%%%%%%%%%%%%%%%%%%%%%%%%%%%%%%%%%%%%%
\section{Model and Scales}\label{sec:model}

We focus on the attractive regime of the unpolarized {\it Gaudin}-{\it Yang}
model~\cite{PhysRevLett.19.1312,*GAUDIN196755}
{in a finite box with periodic boundary}
conditions which, in first quantization, is given by the Hamiltonian
\beq
\label{Eq:H}
\hat H = -\frac{\hbar^2}{2 m}\sum_{i=1}^{N}\frac{\partial^2}{\partial x_i^2} - g\,\sum_{i<j}{\delta(x^{}_i-x^{}_j)}\,.
\eeq
The coupling~$g$ is related to the $s$-wave scattering length~$a_{s}$,~$g\sim 1/a_{\rm s}$,
see, e.g., Ref.~\cite{2000EJPh21.435B}.
{We use conventions such that~$g >0$ corresponds to an
attractive {interaction and work} in units where $k^{}_\textnormal{B} = \hbar = m = 1$ with $m$ being the
mass of the fermions, equal for both spins. As previously mentioned, our
attention is restricted to {the case with two fermion species interacting via a contact interaction, an}
example for a \emph{Luttinger liquid}~\cite{1963JMP.....4.1154L}.}
For our ground-state calculations, we employ the
techniques previously used {in Refs.~\cite{GCS1D, ImprovedActionsDrut, GCS2D}.  Specifically, we} formulate the
given quantum many-body problem on a discretized Euclidean spacetime of dimensionless extent $N^{}_x \times N^{}_\tau$.
Using a symmetric Trotter-Suzuki decomposition followed by an auxiliary field transformation, we arrive at path-integral expressions for our desired observables,
which we evaluate via the {Hybrid Monte} Carlo (HMC) algorithm.

{The calculations presented in this work have been}
carried out by projecting the ground state out of a guess wave function of fixed particle number $N = N^{}_{\uparrow} + N^{}_{\downarrow}$.
This even integer along with the ring circumference $L = N^{}_{x} \ell$ with lattice spacing $\ell$ and the attractive coupling {strength $g > 0 $ comprise}
the physical input parameters where only the latter two are dimensionful. As is typical in 1D ground-state studies,
from these two quantities we define a single intensive {dimensionless parameter $\gamma= g/n$,
where $n = N/L$ is} the particle-number density. {The extent of the imaginary time direction is $\beta = \tau N^{}_{\tau}$,
defining $\tau$ as the temporal lattice spacing.  Therefore an extrapolation to the large $\beta \veps^{}_{\text{F}}$ limit is required,
where $\veps^{}_{\text{F}} = k^{2}_{\text{F}}/(2 m)$ and $k^{}_{\text{F}} = \pi n/2$.
Note that, in all cases, we have fixed the (spatial) lattice spacing to unity, {which sets the}
length and momentum scales.}

%%%%%%%%%%%%%%%%%%%%%%%%%%%%%%%%%%%%%%%%%%%%%
\section{Results}\label{sec:res}

{In this section,  we
discuss our results for all previously mentioned observables
as a function of the dimensionless coupling $\gamma$.
In general, we} took $N_x^{} = 80$ lattice sites, consistent with previous
studies for the 1D ground
state~\cite{GCS1D}. Results were obtained by averaging $\sim 5000$ decorrelated samples{. Typical autocorrelation times for the total energy are of the order of $10^{-2}$ and the sampling frequency of order $1$, to ensure decorrelation also for other quantities under study. Unless otherwise noted, error estimates were}
obtained from statistical uncertainties by considering the standard deviation of the mean.}
For a discussion of lattice size and discretization effects, we refer the reader to the Appendix.

%%%%%%%%%%%%%%%%%%%%%%%%%%%%%%
\subsection{Ground-state energy}\label{subsec:gse}
\begin{figure}[t]
  \includegraphics[width=\columnwidth]{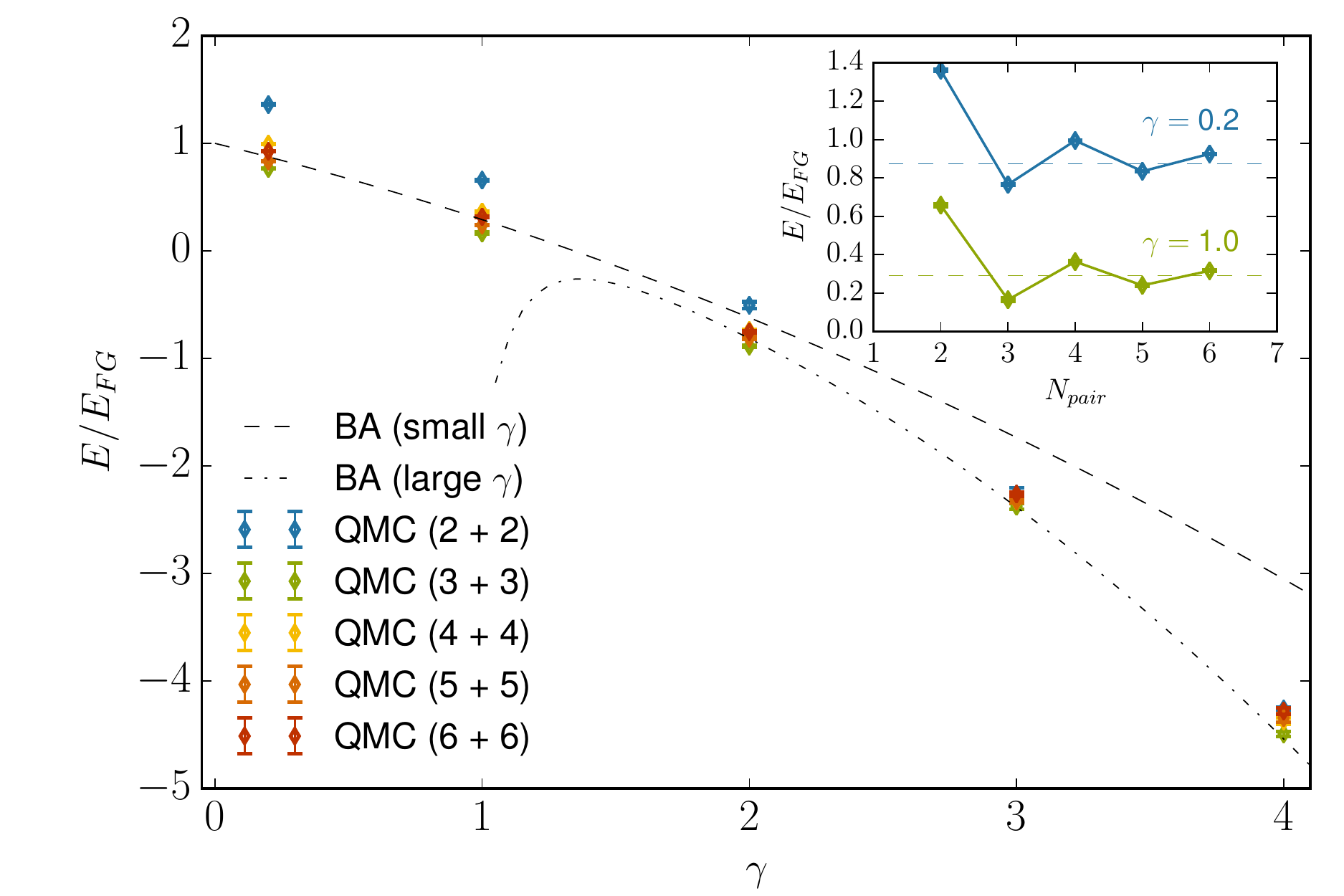}
  \caption{Equation of state for $N = 4$ to $12$ particles as a function of $\gamma$,
  extrapolated {to the infinite-volume limit}. The dashed and dot-dashed lines
  are the results from the {\it Bethe}-ansatz (BA)~\cite{wadati} {in the thermodynamic limit} for weak and strong coupling,
  respectively. We observe fast convergence to the thermodynamic limit.}.
  \label{Fig:EOS}
\end{figure}

{
As a first cross-check, we re-computed the ground-state energy
as a function of the coupling strength and particle number.

In Fig.~\ref{Fig:EOS},
we compare our results for~$E/E_{\rm F}$ with the weak-coupling expansion,
\be
\frac{E}{E_{\rm F}}= 1 - \frac{6\gamma}{\pi^2} - \frac{\gamma^2}{\pi^2} + \dots\,,
\ee
and the strong-coupling expansion,
\be
\!\!\!\!\!\!\frac{E}{E_{\rm F}} = -\frac{3}{\pi^2}\gamma^2
+ \left(\frac{\gamma}{1\!-\!2\gamma}\right)^2\left( 1 \!+\! \frac{4\pi^2}{15(1\!-\!2\gamma)^3}\right)
+\dots\,,\label{eq:scexp}
\ee
in the thermodynamic limit as obtained from the {\it Bethe}-ansatz~\cite{wadati, tracyWidom},
where $E_{\rm F}/L=\kf^3/(3\pi)=(N/L)^3\pi^2/24$ is the ground-state energy of the non-interacting two-component
{\it Fermi} gas.
We observe that our results are in excellent agreement
with the weak-coupling expansion for $\gamma \lesssim 2$ and with
the strong-coupling expansion for  $\gamma \gtrsim 2$.
Moreover, the thermodynamic limit appears to be approached rather rapidly,
see also the inset of Fig.~\ref{Fig:EOS}.

The exact (binding) energy of one spin-up and one spin-down fermion
interacting via a contact interaction in the infinite-volume limit is given by~$E_{1+1}=-g^2/4$~\cite{Griffiths},
corresponding to~$E_{1+1}/E_{\rm F}=- 3\gamma^2/\pi^2$.
Thus, we observe that the ground-state energy per pair is simply
given by the energy of the $1+1$-body problem,~$E/N_{\text{pairs}}=-g^2/4$, at leading order in the strong-coupling
expansion. Loosely speaking, the dynamics in the
strong-coupling limit may therefore be viewed as dominated
by the formation of tightly bound pairs built up from one spin-up and one spin-down fermion.
In the strict infinite-coupling limit~$1/\gamma =0$,
the many-body system {can be viewed as a gas of composite bosons, the} so-called {\it Tonks}-{\it Girardeau} gas~\cite{tggas}.

In the weak-coupling limit, a {finite
gap~$\Delta/E_{\rm F} \sim |\gamma|{\rm e}^{-\pi^2/(2|\gamma|)}$ has} been found to emerge
between the singlet
ground state and the first triplet excited state together with
gapless density fluctuations~\cite{PhysRev.130.1605,*PhysRev.130.1616,Krivnov,2004PhRvL..93i0408F,wadati}. Consequently,
the dynamics of the many-body system in this limit is
associated with a {\it Bardeen}-{\it Cooper}-{\it Schrieffer} (BCS) superfluid.
For a detailed discussion of the many-body phase diagram, we refer the reader to Ref.~\cite{2004PhRvL..93i0408F}.
Here, {we
only highlight that a smooth crossover from the formation of tight bosonic molecules in the limit~$1/\gamma\to -\infty$
to {\it Cooper pairing} in the limit~$1/\gamma \to \infty$
is found to occur at~$\gamma\sim 2$ in this system.} At this point, the size of the bosonic pairs is
of the order of the distance between the fermions~\cite{2004PhRvL..93i0408F,wadati}.
Indeed, for the two-body problem in the infinite-volume limit,
the ``diameter"~$d_0$ of the bosonic pair\footnote{We define the diameter~$d_0$ as $|\Phi(0,d_0/2)|^2=|\Phi(0,0)|^2/{\rm e}$,
where~$\Phi(x_{\uparrow},x_{\downarrow})$ is the ground-state wave function. Note that~$\Phi$ is only
a function of~$|x_{\uparrow}-x_{\downarrow}|$ in the infinite-volume limit.}
is given by~$d_0=2/g$, see, e.g., Ref.~\cite{Griffiths}. Thus, we have~$d_0n=2/\gamma$ which
may be viewed as a measure for the crossover point in terms of the coupling at which
the properties of the system change significantly.

In the following subsections we do not aim at a detailed quantitative discussion of the phase diagram
but focus on our results for the
momentum distribution, the one- and two-body density matrices, and the
particle partition entanglement entropy from few to many fermions.}
%

%%%%%%%%%%%%%%%%%%%%%%%%%%%%%%%%%
\subsection{One-body density matrix and momentum distribution}

{The one-body density matrix~$\rho_1^{(\sigma)}$ in principle allows {us} to compute all
single-particle expectation values and
is defined as ground-state expectation value of
a creation and annihilation operator:
\be
\rho_1^{(\sigma)}(x,x^{\prime})=\langle \hat{\psi}^{\dagger}_{\sigma}(x) \hat{\psi}_{\sigma}(x^{\prime})\rangle\,,
\ee
where~$\sigma$ refers to the spin index and
the operators~{$\hat{\psi}_{\sigma}\ (\hat{\psi}_{\sigma}^{\dagger})$} denote annihilation (creation) operators.
In terms of a general $N$-body wave
function~$\Phi(x_{\uparrow,1},x_{\downarrow,1},\dots,x_{\uparrow,N_{\uparrow}},x_{\downarrow,N_{\downarrow}})$,
the one-body density matrix associated with, e.g., the spin-up fermions is given by
\be
&& \rho_1^{(\uparrow)}(x,x^{\prime})\!=\!
N_{\uparrow}\!\!\int_{-\frac{L}{2}}^{\frac{L}{2}}\!{\rm d}y_{2}\;\cdots\!\!\int_{-\frac{L}{2}}^{\frac{L}{2}}\! {\rm d}y_{N}\Phi^{\ast}(x,y_2,\dots,y_N)\nn\\
&& \qquad\qquad\qquad\qquad\qquad\qquad\times\; \Phi(x^{\prime},y_2,\dots,y_N)\,,
\ee
and correspondingly for the spin-down fermions. The (single-particle) momentum distribution~$\tilde{n}_{k,k^{\prime}}^{(\sigma)}$,
i.e. the Fourier transform of
the one-body density matrix, is implicitly defined as
\be
\rho_1^{(\sigma)}(x,x^{\prime})\!=\!
\sum_{l,l^{\prime}} \varphi_l^{\ast}(x)\tilde{n}_{l,l^{\prime}}^{(\sigma)}\varphi_{l^{\prime}}(x^{\prime})\,,
\label{eq:momdistdef}
\ee
where
\be
\varphi_l(x)=\frac{1}{\sqrt{L}}{\rm e}^{{\rm i}\omega_l x}
\ee
and~$\omega_l = 2\pi l/L$ for the periodic box of extent~$L$ considered in this
work.

The one-body density matrix~$\rho_1^{(\sigma)}$ determines the overlap of a state,
in which a fermion with spin~$\sigma$ has been removed
from the ground state at point~$x^{\prime}$, with a state, in which a fermion with the same spin~$\sigma$ has
been removed at point~$x$.
Correspondingly, the associated single-particle momentum distribution
determines the overlap of a state,
in which a fermion with spin~$\sigma$ and momentum~$k^{\prime}$ has been removed
from the ground state, with a state, in which a fermion with the same spin~$\sigma$ but momentum~$k$ has
been removed. From the definition of the single-particle momentum distribution,
it follows immediately that it
is only finite for~$|k|\leq \kf$ and~$|k^{\prime}|\leq \kf$ in the non-interacting limit.
\begin{figure}[t]
 \includegraphics[width=0.85\columnwidth]{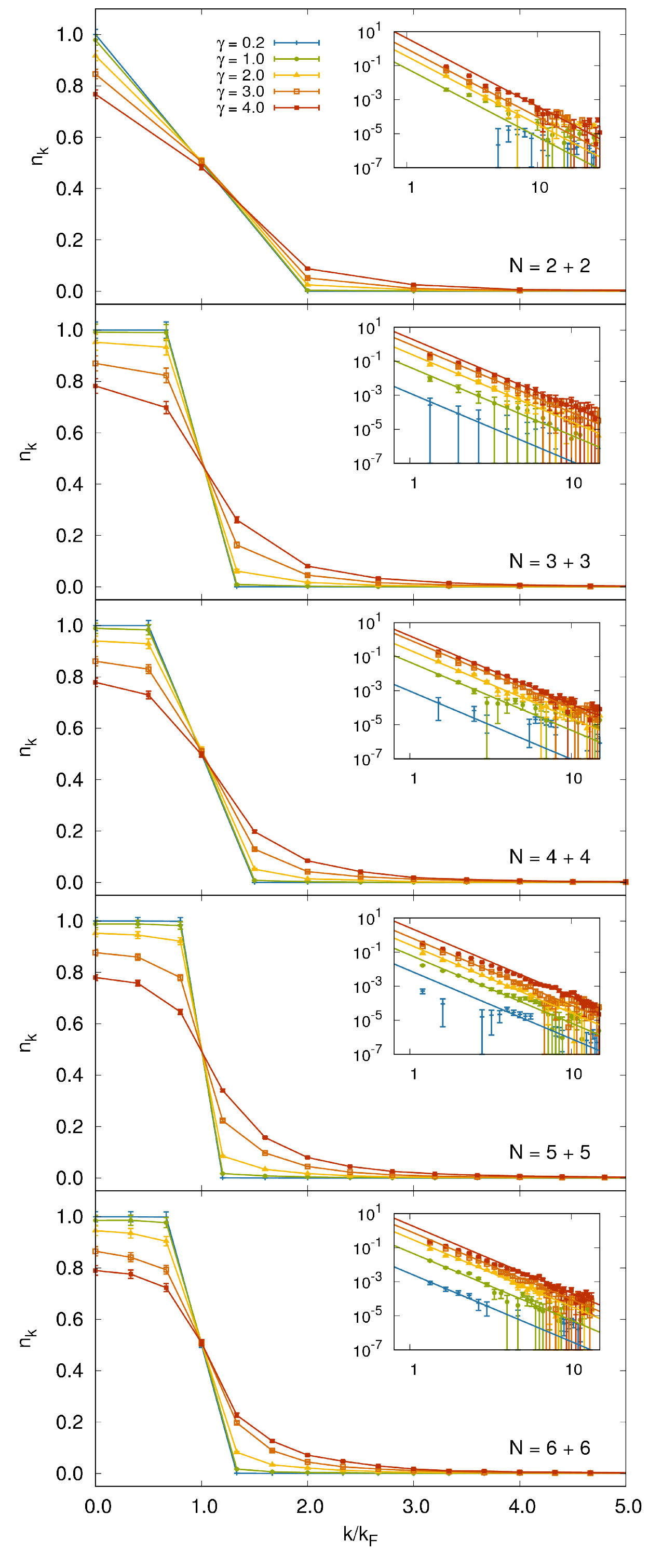}
 \caption{Diagonal part of the momentum distribution $n_{k}$
 as a function of $k/k^{}_F$ for various~$N$. The insets show the
 asymptotic behavior on a double logarithmic scale,
 where solid lines represent linear fits of the data.}
 \label{Fig:MomDist}
\end{figure}

In a periodic box, the one-body density matrix of the non-interacting system
can be computed analytically. We find
\be
&& \rho_1^{(\sigma)}(x,x^{\prime}) = \frac{1}{L}\Bigg( 1 + 2\sum_{j=1}^{\bar{N}_{\sigma}}\cos\left(\omega_j(x-x^{\prime})\right) \nn\\
&& \qquad\qquad\qquad\qquad +\, \delta_{(N_{\sigma}\text{mod}\, 2), 0} \Big[ \cos \left(\omega_{\bar{N}_{\sigma}+1}(x-x^{\prime})\right) \nn\\
&&  \qquad\qquad\qquad\qquad\qquad - \cos \left(\omega_{\bar{N}_{\sigma}+1}(x+x^{\prime})\right)
\Big] \Bigg)\,,\label{eq:obdm}
\ee
where $\bar{N}_{\sigma}=(N_{\sigma}-1)/2$ for odd~$N_{\sigma}$ and~$\bar{N}_{\sigma}=(N_{\sigma}-2)/2$ for
even~$N_{\sigma}$.

For odd~$N_{\sigma}$, we observe that the one-body density matrix
of the non-interacting system is a translation-invariant quantity as it only
depends on the distance between~$x$ and~$x^{\prime}$. For even~$N_{\sigma}$, however,
the one-body density matrix of the non-interacting system is no longer
translation-invariant in a periodic box but depends on~$x$ and~$x^{\prime}$ separately, see Eq.~\eqref{eq:obdm}.
Nevertheless, in the large-$N_{\sigma}$ limit, the term breaking translation invariance is only subdominant, implying
that the one-body density matrix becomes a translation-invariant quantity in the thermodynamic
limit, as it should be. In fact, we have
\be
 \rho_1^{(\sigma)}(x,x^{\prime}) = \frac{\sin \left(\pi n_{\sigma} |x-x^{\prime}|\right)}{\pi |x-x^{\prime}|}\,
 \label{eq:rho1free}
\ee
{for fixed~$n_{\sigma}=N_{\sigma}/L$ with~$N_{\sigma}\to\infty$ and~$L\to \infty$}.

We emphasize that the breaking of translation invariance in systems with even~$N_{\sigma}$ is a direct consequence
of the fact that the ground-state wave function of the non-interacting system is not
an eigenstate of the center-of-mass momentum operator $\hat P_\text{tot}$; it is, however, an eigenstate of $\hat P^2_\text{tot}$.
For odd~$N_{\sigma}$, on the other hand, the ground-state wave function
{\it is} an eigenstate of $\hat P_\text{tot}$ with zero {eigenvalue.\footnote{As the Hamiltonian and
parity operators commute with each other, the ground-state wave function (including the center-of-mass motion) can be chosen to be
an eigenstate of the parity operator.
Note that the part of the ground-state wave function describing the relative motion of the fermions has even parity whereas the parity
of the center-of-mass wave function can be chosen at will. In our numerical implementation, conventions effectively correspond
to choosing the center-of-mass wave function to have odd parity for even~$N_{\sigma}$. For odd~$N_{\sigma}$,
we choose the center-of-mass wave function to have odd parity if $(N_{\sigma}-1)/2$ is odd, and otherwise even.}}
Since the ground-state
wave function of the fully interacting system is effectively generated by exciting the ground-state
wave function of the non-interacting system according to the momentum-conserving interaction,
we conclude that translation invariance of the ground-state wave function
is preserved in our QMC studies for systems with odd~$N_{\sigma}$ but is violated
for systems with even~$N_{\sigma}$, see also Ref.~\cite{Kemler:2016wci} for a
discussion of this issue for systems in
(anti)periodic boxes.
We return to this below when discussing
our results for the one-body density matrix.

From the one-body density matrix in Eq.~\eqref{eq:obdm}, the
momentum distribution~$\tilde{n}_{l,l^{\prime}}$ of the non-interacting system is readily obtained. We find
\be
\tilde{n}^{(\sigma)}_{l,l^{\prime}} &=& \delta_{l,l^{\prime}}\theta( \bar{N}_{\sigma} - |l| ) \nn\\
&&  \quad + \frac{1}{2}\delta_{(N_{\sigma}\text{mod}\, 2), 0}\left( \delta_{l,(\bar{N}_{\sigma}+1)} - \delta_{l,-(\bar{N}_{\sigma}+1)}\right)\nn\\
&& \qquad\qquad\;\times\,
 \left( \delta_{l^{\prime},(\bar{N}_{\sigma}+1)} - \delta_{l^{\prime},-(\bar{N}_{\sigma}+1)}\right)\,,
 \label{eq:nkkni}
\ee
where~$\theta(x)=1$ for~$x\geq 0$ and~$\theta(x)=0$ otherwise.
\begin{figure}[t]
  \includegraphics[width=\columnwidth]{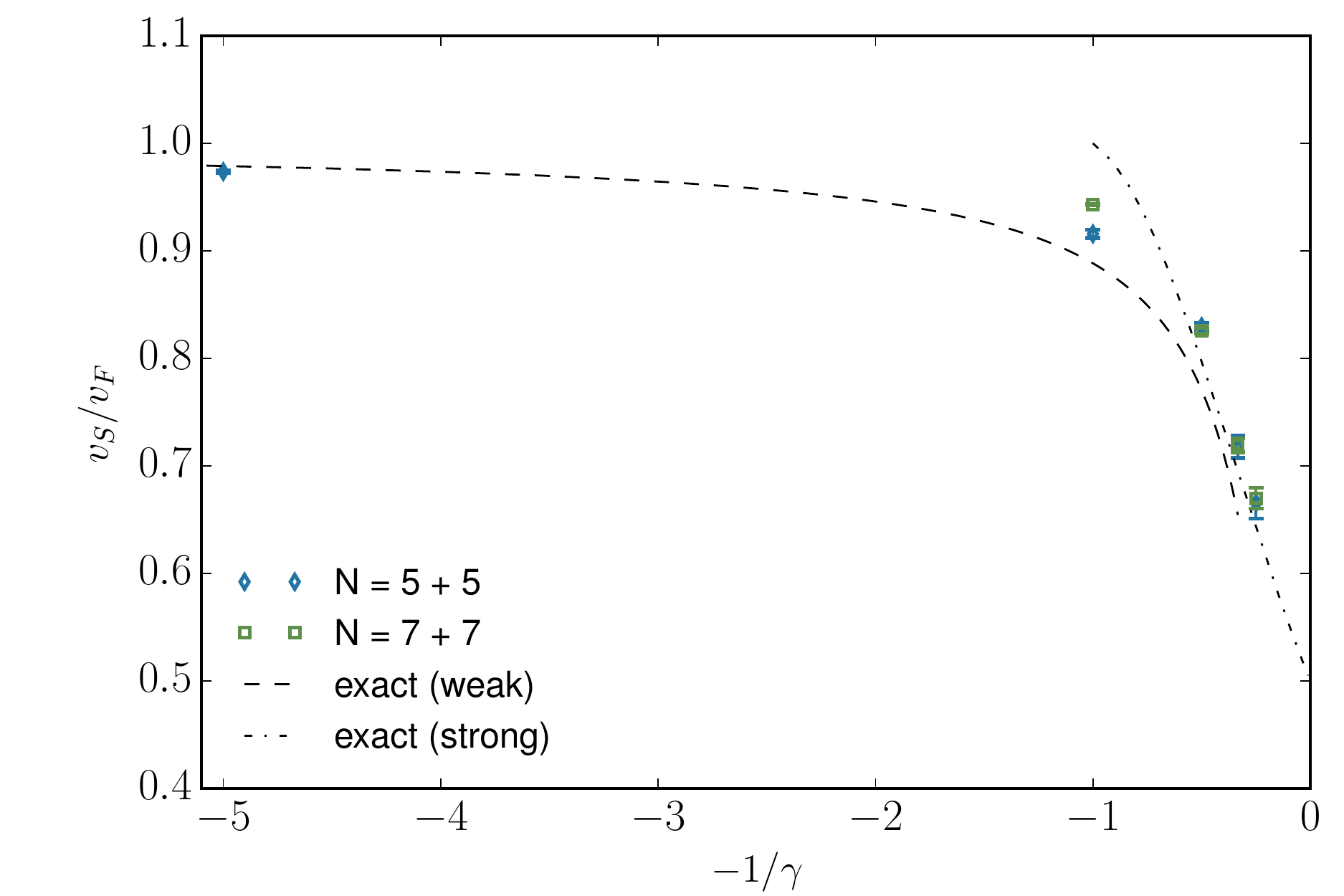}
  \caption{Estimates for the sound velocity~$v_{\rm s}/v_{\rm F}$ as a function of the inverse
  {coupling, where~$v_{\rm F}$ is} the {\it Fermi} velocity. {Errorbars reflect propagated uncertainties in $\eta$ from a fit of the form \eqref{eq:nlowk} to the momentum distribution at low momenta.}}
  \label{Fig:sound}
\end{figure}

In Fig.~\ref{Fig:MomDist}, we show our results for the diagonal part of the momentum
distribution $n_{k}\equiv \tilde{n}^{(\uparrow)}_{l,l}= \tilde{n}^{(\downarrow)}_{l,l}$
as a function of $k/\kf$ for various particle numbers and coupling strengths~$\gamma$.
For small values of the coupling, $0<\gamma \lesssim 1$, we observe that the momentum distribution
is still well described by the non-interacting momentum distribution given in Eq.~\eqref{eq:nkkni},
independent of the total particle number~$N$. For stronger couplings, $\gamma \gtrsim 2$, the system is then dominated
by the formation of tightly bound dimers where the crossover to this regime from the
weakly coupled regime dominated by {\it Cooper} pairing occurs at~$\gamma\sim 2$,
see our discussion in Sec.~\ref{subsec:gse}. In the regime associated with~$\gamma\gtrsim 2$,
the momentum distributions clearly deviate from their non-interacting counterparts.
More specifically,
even states with very low momenta are now excited above the
{\it Fermi} point~$\kf$. Loosely speaking, the momentum distributions
{effectively start to flatten out} when the coupling is increased beyond~$\gamma \sim 1$
and therefore
these distributions lose their characteristic feature of a sharp drop present in the weak-coupling limit.

In order to further quantify the change in the momentum distributions when the coupling is increased,
we analyze its scaling behavior close to the {\it Fermi} {point~$\kf$.}
For spin-balanced systems, $n=2n_{\uparrow}=2n_{\downarrow}$, and $n|x-x^{\prime}| \gg 1$,
the one-body density matrix  in the thermodynamic limit is known to scale
as follows~\cite{PhysRevB.9.2911,PhysRevLett.47.1840,2004JPhB...37S...1C}:
\be
\rho_1 (x,x^{\prime}) \equiv \rho_1^{(\sigma)}(x,x^{\prime})
 \sim \frac{\sin\left(\pi n|x-x^{\prime}|\right)}{(n|x-x^{\prime}|)^{\frac{1}{\eta}+\frac{\eta}{4}}}\,,
\label{eq:obdmscal}
\ee
where~$n=2\kf/\pi$.
A comparison with the exact solution~\eqref{eq:rho1free} for the free gas (i.e. $\gamma=0$)
immediately yields~$\eta=2$.

From Eq.~\eqref{eq:obdmscal}, we obtain
the scaling behavior of the single-particle {momentum distribution
(close) below the {\it Fermi} point~$\kf$ in}
the thermodynamic limit, see also~Ref.~\cite{RevModPhys.80.1215}:
\be
n(k) \sim (\kf -|k|)^{\frac{1}{\eta}+\frac{\eta}{4}-1}\,.\label{eq:nlowk}
\ee
The scaling exponent~$\eta$ is directly related to the sound velocity~$v_{\rm s}$ of density fluctuations
in our 1D {\it Fermi} gas. Indeed, we have~$v_{\rm s}/v_{\rm F} = 2/\eta$
with~$v_{\rm F}=\kf$ being the {\it Fermi} velocity~\cite{PhysRevB.9.2911,PhysRevLett.47.1840}.
With~$\eta=2$, we find~$v_{\rm s}/v_{\rm F}=1$
for the free gas as expected.

{For weak attractive interactions~\cite{Krivnov},~$\gamma\to 0^{+}$, the} sound velocity is given by
\be
\frac{v_{\rm s}}{v_{\rm F}}=1- \frac{\gamma}{\pi^2} + \frac{\gamma^2}{2\pi^4}\left(\ln|\gamma|-2\right) + \dots\,,
\ee
whereas it reads
\be
\frac{v_{\rm s}}{v_{\rm F}}= \frac{1}{2} + \frac{1}{2\gamma} + \frac{3}{4\gamma^2} - \frac{3}{4\gamma^3}
+\dots\,
\ee
in the limit~$1/\gamma\to 0$, i.e. in the strong-coupling limit~\cite{PhysRev.130.1605}.
Assuming that the sound velocity is a monotonic function of the coupling~$\gamma$, we conclude from
these expansions
that~$\eta$ varies between~$\eta=2$ at~$\gamma=0$ and~$\eta=4$ in the limit~$1/\gamma\to 0$.
Recall our conventions
for the sign of the coupling~$\gamma$, see~Eq.~\eqref{Eq:H}.

In this work, we exploit the scaling law~\eqref{eq:nlowk} to estimate the sound velocity
from a fit of our numerical data
in the low-momentum {regime~$k\lesssim \kf$ to the ansatz~$\xi_0  (\kf-|k|)^{\xi_1}$
based on the two parameters~$\xi_0$ and~$\xi_1$.}
Of course, a high-precision determination of the low-momentum scaling behavior
and the associated sound velocity requires to study larger
particle numbers and even larger box sizes than considered in our present work in order
to push the system closer to the thermodynamic limit.\footnote{In principle, the speed of sound
can also be computed directly from the derivative of the chemical potential with respect to the density.
However, the computation of the chemical potential defined as a derivative of the ground-state energy with respect
to the particle number requires the computation of the ground-state energy of
spin-imbalanced systems which is beyond the scope of the present work.}
Still, our present results for the
sound velocity obtained from such a fit procedure
already appear to be in reasonable agreement with the
existing results for this quantity~\cite{2004PhRvL..93i0408F}, in particular
with the weak- and strong coupling expansion given above, see Fig.~\ref{Fig:sound}. More specifically,
we observe that the sound velocity remains close to the {\it Fermi} velocity for~$\gamma \lesssim 2$.
For~$\gamma \gtrsim 2$, the sound velocity then starts to decrease rapidly, suggesting that
the systems enters the crossover regime between the ``phase" dominated by
 {\it Cooper} pairing
at small attractive couplings to a ``phase" governed by the formation of a
gas of tightly bound bosonic molecules, in accordance with earlier studies~\cite{2004PhRvL..93i0408F}.
\begin{table}[t]
\begin{tabularx}{\columnwidth}{@{\extracolsep{\fill}} c c | c c  }
\hline \hline
$N$	& $\gamma$	& $C/(L k_{\textnormal{F}}^{4}) $ (this work)	& $C/(L k_{\textnormal{F}}^{4})$ (Ref. \cite{GCS1D})	\\
\hline
2+2	& 0.2	& 0.003(1)	& 0.00204(3)	\\
	& 1.0	& 0.02(2)	& 0.063(2)	\\
	& 2.0	& 0.36(2)	& 0.35(1)	\\
	& 3.0	& 1.01(2)	& 1.03(3)	\\
	& 4.0	& 2.3(3)	& 2.36(2)	\\
\hline
4+4	& 0.2	& 0.0013(2)	& 0.00182(5)	\\
	& 1.0	& 0.059(6)	& 0.0582(6)	\\
	& 2.0	& 0.330(5)	& 0.324(4)	\\
	& 3.0	& 0.99(3)	& 0.99(1)	\\
	& 4.0	& 2.17(9)	& 2.24(2)	\\
\hline
6+6	& 0.2	& 0.0012(4)	& 0.00178(3)	\\
	& 1.0	& 0.055(2)	& 0.0563(6)	\\
	& 2.0	& 0.33(1)	& 0.311(4)	\\
	& 3.0	& 0.94(2)	& 0.94(1)	\\
	& 4.0	& 2.20(8)	& 2.14(2)	\\
\hline  \hline
\end{tabularx}
\caption{\label{Table:Contact}
{Estimates for the contact density $C/(L\kf^{4})$ for different values of the dimensionless
coupling $\gamma$ and the total particle number $N=N_{\uparrow}+N_{\downarrow}$ as obtained from linear fits
to the large-momentum tails of the momentum distributions on a double-logarithmic scale as
presented in the insets of Fig.~\ref{Fig:MomDist}.}
}
\end{table}

We note that, as in the case of the ground-state energy, a fast convergence of our results
to the thermodynamic limit is observed. In general,
however, the convergence is faster for odd~$N_{\sigma}$ as terms violating translation invariance
are absent in this case, see Eq.~\eqref{eq:obdm} and our discussion below.}

{Let us now turn to the scaling behavior of the momentum distribution
at high momenta, which determines {\it Tan}'s contact density $C /(L\,\kf^{4})$.\footnote{A detailed
discussion of {\it Tan}'s relations for 1D systems can be found
in Ref.~\cite{2011AnPhy.326.2544B}.}
For the latter, we extracted
estimates from the asymptotic behavior of
the momentum distribution~\cite{Tan20082952,*Tan20082971,*Tan20082987},
\beq
C \equiv \lim_{|k|\to\infty}k^4 n_k\,,
\eeq
by performing a linear fit of our results for $n_{k}^{}$ for momenta with $|k| > \kf$
on a double-logarithmic scale, see also insets of Fig.~\ref{Fig:MomDist}.
{In our fits, we have only taken
results for momenta with $|k| > 2\kf$ into account. Above this scale, we observe that the fits only exhibit a
weak dependence on the actual fit range. { Note that, for very dilute systems (i.e. small particle numbers) the high momentum part is subject to noise which explains the seemingly odd behavior at large momenta.}}

{\it A priori}, it is not evident where the above large-momentum asymptotics sets in.
It was found in previous studies in 3D at unitarity~\cite{momDecayCutoff} that
the onset scale is close to $k/\kf \sim 2$. In a QMC study of the corresponding 2D system~\cite{GCS2D},
it was observed that the onset scale increases with the coupling strength.
At least in 1D, a rough estimate for this onset scale
may be obtained by comparing the ``diameter"~$d=2/(\gamma n)$ of the bosonic
bound state associated with the two-fermion problem with the
{\it de Broglie} wavelength of a given fermion~$\lambda=2\pi/k$.
The typical momentum of a fermion within the bound state may be estimated to be of
the order of~$k_0=2\pi/d$. This momentum scale should be
compared to the momentum~$k$ of a given fermion.
If~$k\lesssim k_0=2\gamma\kf$, then the fermion is not sensitive to the
details of the
short-range physics of our system but only to the long-range aspects.
The long-range physics in our model is indeed
immediately affected by an increase of the coupling,
as indicated by our study of the scaling exponent~$\eta$ of the single-particle
momentum distribution determining the decay
of the one-body density matrix in the long-range limit.
If~$k \gtrsim k_0=2\gamma\kf$, then the fermion is sensitive to the details of the
short-range physics of our 1D system. The latter case is associated with the dynamics which, e.g.,
determines {\it Tan}'s contact density.

In accordance with this simple argument,
we indeed find in our numerical studies that the $k^{-4}$ decay law sets in at higher scales~$k/\kf$ when
the coupling~$\gamma$ is increased for a fixed particle number, see also the insets of Fig.~\ref{Fig:MomDist}.}
{In {Tab.~\ref{Table:Contact}, results} for
the contact density for various couplings} and particle numbers are provided. We note agreement with results previously
obtained in Ref.~\cite{GCS1D} {for even~$N_{\sigma}$}
using a different definition of {\it Tan}'s contact along with the {\it Feynman}-{\it Hellmann} theorem.
\begin{figure*}[t]
 \includegraphics[width=2.05\columnwidth]{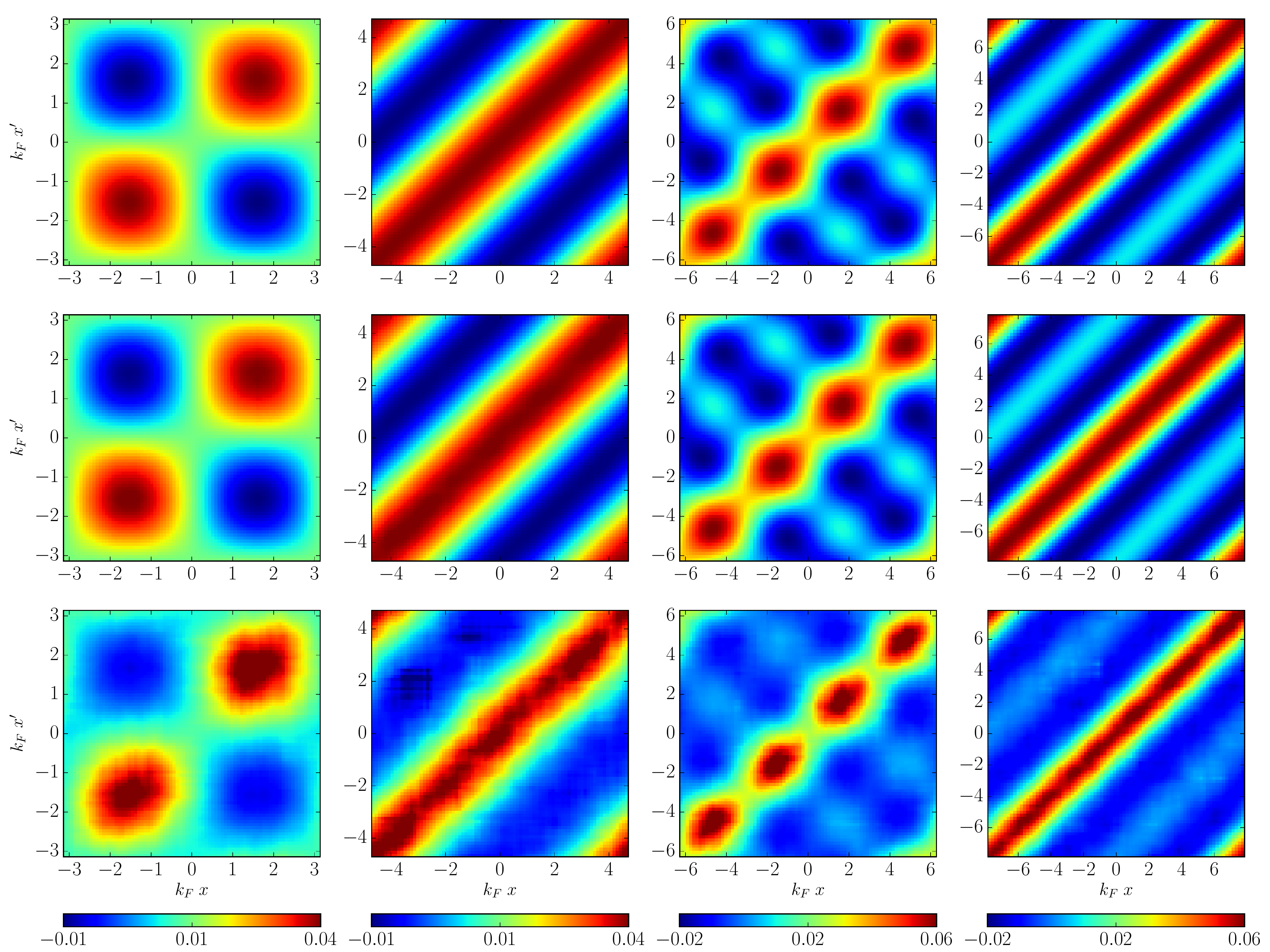}
 \caption{\label{Fig:obdm}{One-body density matrix $\rho_1(x,x^{\prime})$
 as a function of the dimensionless coordinates~$\kf x$ and~$\kf x^{\prime}$
 for $N=2+2, 3+3, 4+4, 5+5$ fermions (from left to right) and~$\gamma=0, 0.2, 3.0$ (from top to bottom)
 in a periodic box with extent~$\kf L$ where~$L$ is fixed. The analytic result for the non-interacting limit ($\gamma=0$)
 is given in Eq.~\eqref{eq:obdm}. The color coding is associated with the value of the one-body
 density matrix $\rho_1(x,x^{\prime})$. The violation of translation invariance is clearly visible in
 the results for even~$N_{\sigma}$ but is continuously weakened for increasing~$N_{\sigma}$, see main text for a detailed discussion.}}
\end{figure*}

{Finally, we discuss our results for the one-body
density matrix $\rho_1\equiv \rho_1^{(\uparrow)}=\rho_1^{(\downarrow)}$ being nothing but the
Fourier transform of the
momentum distribution~$n_{k,k^{\prime}}\equiv n_{k,k^{\prime}}^{(\uparrow)} = n_{k,k^{\prime}}^{(\downarrow)}$,
see Eq.~\ref{eq:momdistdef}.
In Fig.~\ref{Fig:obdm}, we present our results for~$\rho_{1}$
as a function of the dimensionless coordinates~$\kf x$ and~$\kf x^{\prime}$ in a periodic box
for $N=2+2, 3+3, 4+4, 5+5$ fermions (from left to right) and~$\gamma=0, 0.2, 3.0$ (from top to bottom).
The color coding is associated with the actual value of the one-body
density matrix $\rho_1$ at the point $(x,x^{\prime})$.
The results for finite~$\gamma$ represent numerical
data from our QMC calculations, whereas the result for the non-interacting
system ($\gamma=0$) was obtained analytically, see Eq.~\eqref{eq:obdm}.

The results shown in Fig.~\ref{Fig:obdm} exemplify our findings for other particle numbers.
As suggested by the analytic solution~\eqref{eq:obdm} for the non-interacting limit,
we observe that the number of oscillations at fixed coupling and box size
increases with increasing particle number. The scale
for these oscillations is set by the density.
The main maxima of the one-body density matrix are found along the lines with~$|x-x^{\prime}|=\nu L$,
where~$\nu\in {\mathbb Z}$. However, as already indicated above, we also clearly see that
translation invariance is broken for even~$N_{\sigma}$, whereas it is manifest
for odd~$N_{\sigma}$. This invariance is progressively restored as the particle number is increased.
The mild violation of translation invariance for~$\gamma=3.0$ and odd~$N_{\sigma}$
in Fig.~\ref{Fig:obdm} is due to statistical uncertainties in our QMC calculations at strong couplings.

In Fig.~\ref{Fig:obdm}, we also find that the width of the band associated with the lines of
main maxima at~$|x-x^{\prime}|=\nu L$ is decreased with increasing coupling strength
and the oscillations tend to flatten, leading to an increased localization of the one-body density
matrix. This observation is consistent with the fact that the dynamics
is governed by the formation of tightly bound bosonic molecules in the strong-coupling limit.
Indeed, given our results for the single-particle momentum distribution,
the increased localization of {the one-body density matrix for} increasing coupling strength
does not come unexpected at all. It is, rather,
a direct consequence of the fact that the single-particle momentum distribution is
increasingly smeared out when $\gamma$ is increased for fixed particle number.
Quantitatively, this is measured by the increase of the
critical exponent~$\eta$ associated with the long-range scaling of the
one-body density matrix, see Eq.~\eqref{eq:obdmscal}, when $\gamma$ is increased.

%%%%%%%%%%%%%%%%%%%%%%%%%%%%%%%%%
\subsection{Pair-correlation function}

{In addition to the one-body density matrix, we have calculated the
pair-correlation function, also known as the on-site two-body
density matrix. In one-dimensional systems,
this function has attracted a lot of interest for instance in the {search
for inhomogeneous ground states~\cite{PhysRevA.78.033607}. It} is defined as
\be
\rho_{\rm pair}(x,x^{\prime})&=&\langle \hat{\psi}^{\dagger}_{\uparrow}(x) \hat{\psi}^{\dagger}_{\downarrow}(x)
 \hat{\psi}_{\uparrow}(x^{\prime}) \hat{\psi}_{\downarrow}(x^{\prime})\rangle\,.
\ee
This expression can be rewritten in terms of the ground-state $N$-body wave function~$\Phi$:
\be
&& \rho_{\rm pair}(x,x^{\prime})\!=\! N_{\uparrow}N_{\downarrow}
\!\!\int_{-\frac{L}{2}}^{\frac{L}{2}}\!\!{\rm d}y_{3}\cdots\!\!\int_{-\frac{L}{2}}^{\frac{L}{2}}\!\! {\rm d}y_{N}\Phi^{\ast}(x,x,y_3,\dots,y_N)\nn\\
&& \qquad\qquad\qquad\qquad\qquad\qquad\;\times\; \Phi(x^{\prime},x^{\prime},y_3,\dots,y_N)\,.
\ee
Note that
\be
\int_{-\frac{L}{2}}^{\frac{L}{2}}{\rm d}x\,\rho_{\rm pair}(x,x)= \frac{N_{\uparrow}N_{\downarrow}}{L}\,,\label{eq:npairnorm}
\ee
where~$N_{\uparrow}N_{\downarrow}$ is the number of all possible combinations of one spin-up
fermion with one spin-down fermion in a system with~$N=N_{\uparrow}+N_{\downarrow}$ fermions.

The pair-correlation function determines the overlap of a state,
in which a pair of one spin-up and one spin-down fermion has been removed
from the ground state at point~$x^{\prime}$, with a state, in which such a pair has
been removed at point~$x$.
Correspondingly, the so-called pair-momentum distribution~$\tilde{n}_{k,k^{\prime}}^{(\text{pair})}$,
which is the Fourier transform of
the pair-correlation function, determines the overlap of a state,
in which a pair of one spin-up and one spin-down fermion with momentum~$k^{\prime}$ has been removed
from the ground state, with a state, in which such a pair with momentum~$k$ has
been removed:
\be
\rho_{\rm pair}(x,x^{\prime})\!=\!
\sum_{k,k^{\prime}} \varphi_k^{\ast}(x)\tilde{n}_{k,k^{\prime}}^{(\text{pair})}\varphi_{k^{\prime}}(x^{\prime})\,.
\label{eq:momdistdef}
\ee
From Eq.~\eqref{eq:npairnorm}, it follows immediately that
\be
\sum_k \tilde{n}_{k,k^{\prime}}^{(\text{pair})} = \frac{N_{\uparrow}N_{\downarrow}}{L}\,.
\ee
Note that, by definition, the pair-momentum distribution is related to the propagator of a pair of vanishing size.
\begin{figure}[t]
 \includegraphics[width=\columnwidth]{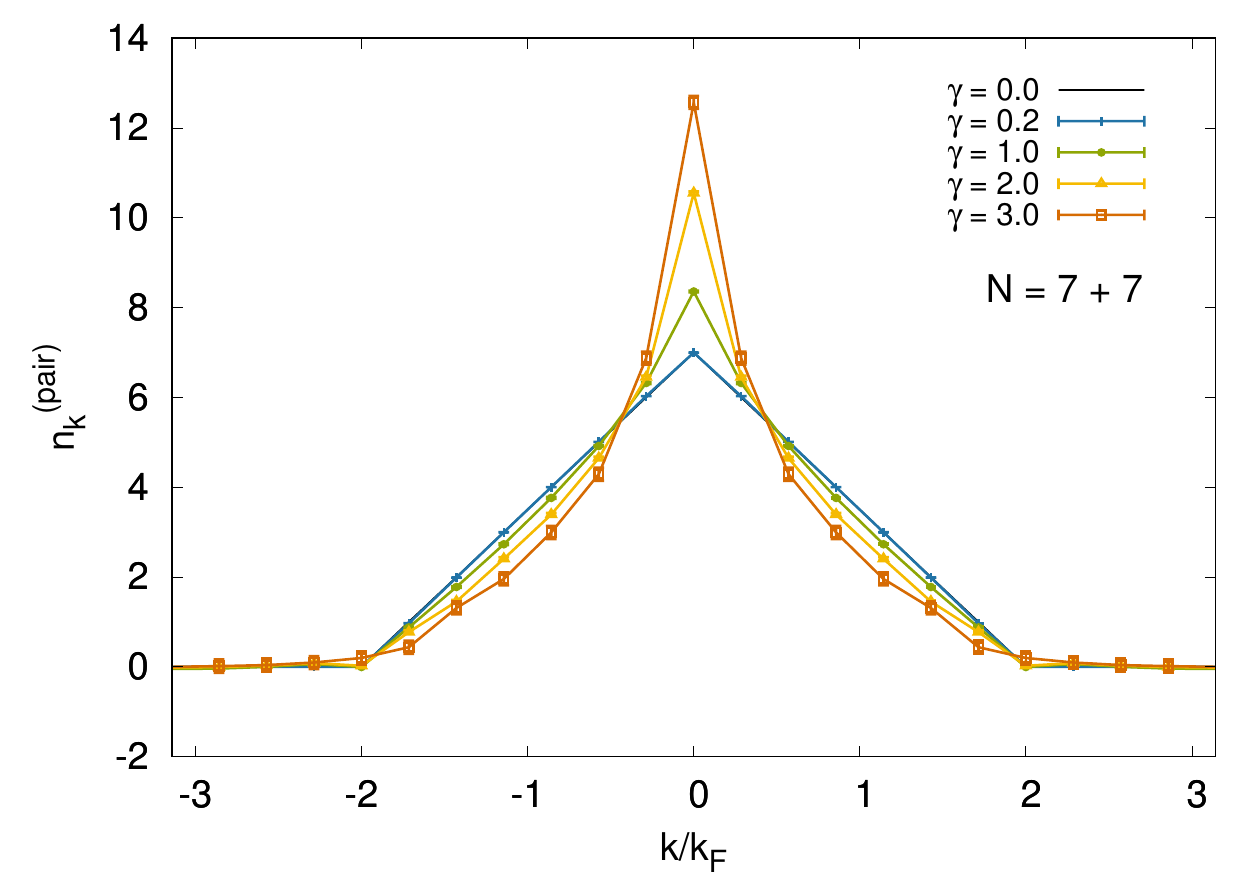}
 \caption{\label{Fig:N2K} {Pair-correlation function
 for~$N=7+7$ fermions and different values of the coupling.}}
\end{figure}

In the non-interacting limit, the pair-correlation function is simply the product of the one-body
density matrices associated with the spin-up and spin-down fermions:
\be
\rho_{\rm pair}(x,x^{\prime})=\rho_1^{(\uparrow)}(x,x^{\prime})\rho_1^{(\downarrow)}(x,x^{\prime})\,.
\ee
We immediately conclude that the pair-correlation function in a periodic box also suffers
from terms violating translation invariance for even~$N_{\sigma}$.
Thus, the convergence to the thermodynamic limit is in general expected to be faster for odd~$N_{\sigma}$.
For our discussion of the pair correlation function in this work, we shall focus on
the latter case from now on.
The associated pair-momentum distribution of the non-interacting system then reads
\be
\tilde{n}_{k,k^{\prime}}^{(\text{pair})}&=&\frac{\delta_{k,k^{\prime}}}{L}\sum_{j=-\infty}^{\infty}
\theta( \bar{N}_{\uparrow}-|j|)\theta(\bar{N}_{\downarrow}-|j+k|)\,.
\ee
We note that~$\tilde{n}_{0,0}^{(\text{pair})}=N_{\downarrow}/L$ for $\bar{N}_{\uparrow} \geq \bar{N}_{\downarrow}$
and vice versa for $\bar{N}_{\uparrow} < \bar{N}_{\downarrow}$.
Without loss of generality, we may now assume $\bar{N}_{\uparrow} \geq \bar{N}_{\downarrow}$ to obtain
\be
\tilde{n}_{k,k^{\prime}}^{(\text{pair})} &=& \frac{\delta_{k,k^{\prime}}}{L} \Big[ (2\bar{N}_{\downarrow}+1)\theta( |\bar{N}_{\uparrow}-\bar{N}_{\downarrow}|-|k|)\nn\\
&& \quad + ( \bar{N}_{\uparrow}+\bar{N}_{\downarrow}+1 - |k|)\theta(|k| - |\bar{N}_{\uparrow}-\bar{N}_{\downarrow}|)\nn\\
&& \qquad\qquad\qquad\quad \times\theta( \bar{N}_{\uparrow}+\bar{N}_{\downarrow}-|k|)\Big]\,.
\label{eq:npair}
\ee
From this expression, we observe that, for spin-balanced systems, the pair-momentum distribution
assumes a global maximum for~$k=k^{\prime}=0$ (see also Fig.~\ref{Fig:N2K}). Phenomenologically,
this implies that removing an on-site pair with zero momentum is most favorable.
This observation is in line with standard BCS theory where pairing of spin-up and spin-down fermions both located on the
{\it Fermi} surface but with opposite momenta is most favorable
 in the presence of an infinitesimally weak but finite attractive coupling, eventually leading to a destabilization of the
{\it Fermi} surfaces.

{We note that, for  spin-imbalanced systems, the pair-momentum distribution
of the non-interacting system remains constant up
to momenta~$Q_{\rm LOFF}\sim |k_{{\rm F},\uparrow}-k_{{\rm F},\downarrow}|$ and then
decreases monotonically, where LOFF refers to {\it Larkin}, {\it Ovchinnikov}, {\it Fulde},
and {\it Ferrell}~\cite{FuldeFerrell64,LarkinOvchinnikov64}.
For interacting spin-imbalanced systems, the pair-momentum distribution}
has even been found to develop maxima at~$\pm Q_{\rm LOFF}$,
see Refs.~\cite{PhysRevA.78.033607,Lee:2011zzo}.
Since~$Q_{\rm LOFF}$ is associated with the center-of-mass momentum of the formed on-site pairs, the
observation of such maxima may
be viewed as a precursor for the formation of LOFF-type ground states, where~$Q_{\rm LOFF}$
sets the scale for
the periodic structure of the ground state in the many-body phase diagram~\cite{FuldeFerrell64,LarkinOvchinnikov64}.
{\it A priori}, however, the mere existence of such maxima in the pair-correlation function
does not necessarily entail that pairs with momenta~$Q_{\rm LOFF}$ describe the lowest-lying two-body
states in the spectrum and that a condensate is formed out of these states, see, e.g., Ref.~\cite{Roscher:2016chy}.
Still, (pronounced) maxima at~$\pm Q_{\rm LOFF}$
may be viewed as an indication that the formation of pairs with momenta~$Q_{\rm LOFF}$ is
favored.

In Fig.~\ref{Fig:N2K}, as a concrete example for the pair-momentum distribution, we show our results
for~$n^{(\text{pair})}_k=\tilde{n}^{(\text{pair})}_{k,k}$ as a function of the momentum~$k$
for a spin-balanced system of $N=7+7$ fermions.
For increasing coupling~$\gamma$, we observe that the pair-momentum distribution progressively narrows,
resulting in an increase of the maximum at vanishing momenta. This may be viewed as an indicator
that pre-formed on-site pairs are favored to occupy the state of zero center-of-mass momentum. Indeed,
we do not expect the {formation of an inhomogeneous (LOFF-type) ground state}
for the spin-balanced {\it Fermi} gas studied
in this work.

Finally we note that the observed {progressive
formation} of a narrow maximum in the momentum distribution associated with
the formation of on-site pairs
is also consistent with the observation that the system is expected to undergo a smooth crossover from {\it Cooper} pairing
at small attractive couplings to a gas of bosonic molecules at~$\gamma \sim 2$, see our discussion above.}
}

%%%%%%%%%%%%%%%%%%%%%%%%%%%%%%%%%
\subsection{Particle-partition entanglement}

Knowledge of the one- and two-body density matrices, as presented above, enables the calculation of the particle-partition entanglement
entropy. In this {section, we show} the evolution, with particle number, of the one-particle bipartite entanglement entropy.
We define the $n$-particle {{\it R{\'e}nyi}} entanglement entropy via
\beq
S_\alpha(n) = \frac{1}{1-\alpha} \ln \tr\left[ \rho_n^\alpha \right],
\eeq
where $\alpha > 1$ is typically an integer but could in principle take any value in between.
The limit $\alpha \to 1$ yields {the {\it von Neumann}}
version of the entanglement entropy:
\beq
S_1(n) = -\tr\left[ \rho_n \ln \rho_n \right]\,.
\eeq
In this work we will focus on $n=1$, as higher-particle density matrices become progressively noisier (and therefore more challenging to
calculate stochastically) as $n$ is increased.
\begin{table}[t]
\begin{tabularx}{\columnwidth}{@{}l *5{>{\centering\arraybackslash}X} X X @{}}
\hline \hline
$\gamma$	& ${\lambda_1^{(1)}}$ & ${\lambda_1^{(2)}}$ \\
\hline
0.2	& 0.0011(9)	& 0.0003(4)	\\
1.0	& 0.06(1)	& 0.022(4)	\\
2.0	& 0.257(9)	& 0.123(7)	\\
3.0	& 0.51(2)	& 0.30(2)	\\
\hline  \hline
\end{tabularx}
\caption{\label{Table:Subleading} {Next-to-leading order coefficient $\lambda_1^{(\alpha)}$ for the {\it von Neumann}
and {\it R\'enyi} entropies for different values of the coupling $\gamma$.}
}
\end{table}

\begin{figure}[t]
\includegraphics[width=\columnwidth]{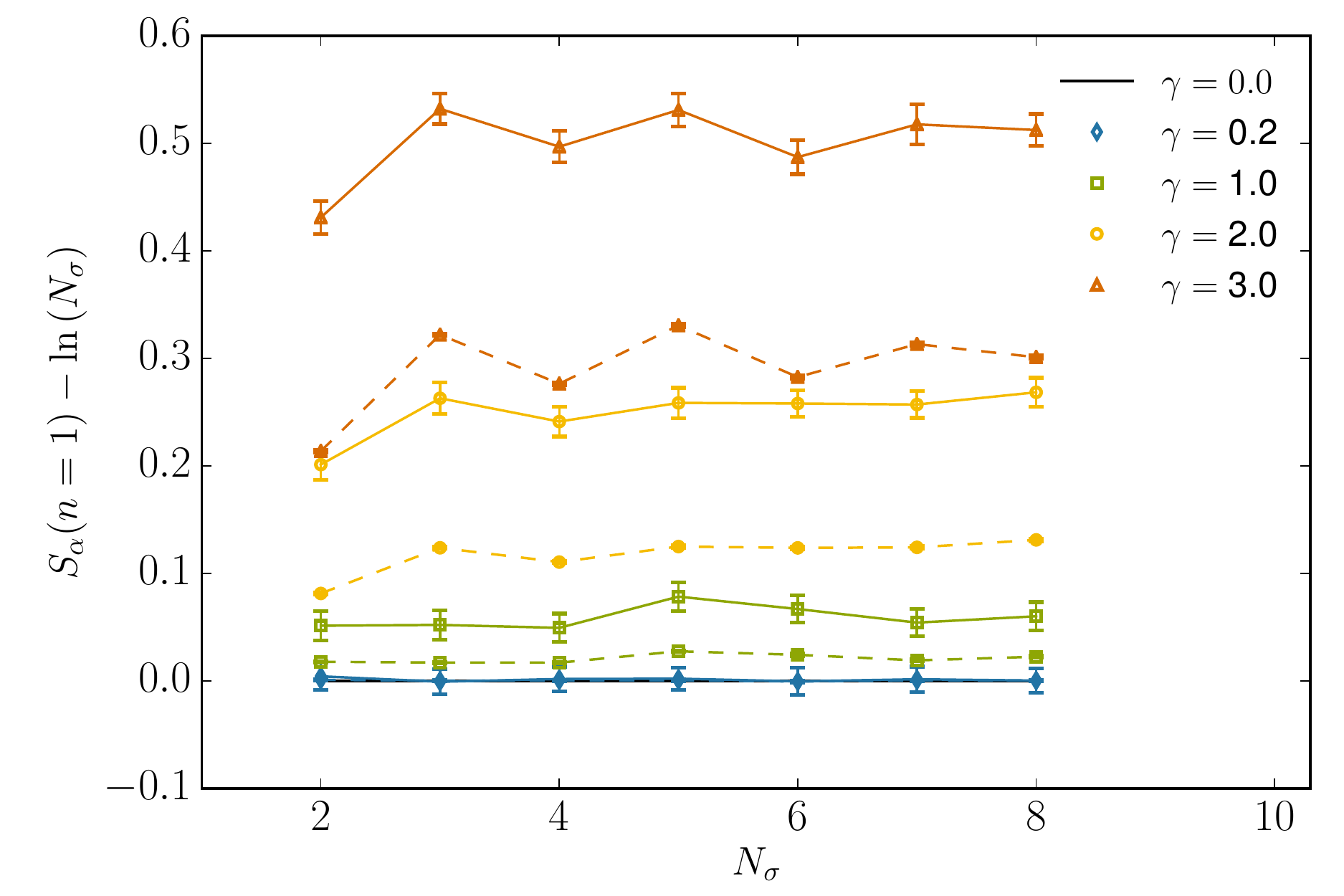}
 \caption{\label{Fig:Salpha} Particle-partition entanglement entropy $S_\alpha$ of 1D fermions
 {as a function of $N_{\sigma}=N/2$
 for} increasing attractive coupling $\gamma$, at fixed partition $n=1$. The solid lines connect the data for {the {\it von Neumann} entropy~$S_1$},
 whereas the dashed lines connect the results for the second {{\it R{\'e}nyi}} entropy $S_2$. Note that $\ln N_{\sigma}$ has been subtracted in order to
 in all cases shows that the $\ln N_{\sigma}$ law is shifted upwards by interaction effects, but otherwise remains valid.
 more clearly display the mild oscillations and the differences between the various data sets. The approximately constant trend of the data
 }
\end{figure}

{In Fig.~\ref{Fig:Salpha}, we show} our results for
{the {\it von Neumann} entropy~$S_1$ and the second {\it R{\'e}nyi}
entropy~$S_2$ at $n=1$ as a function of $N_{\sigma}=N/2$. For} the non-interacting case,
we compare with the answer for spinless fermions~\cite{2017arXiv170310587B,*PhysRevA.78.042326}
and find a similar behavior for very weakly coupled systems.
Our results for the strongly interacting case show mild oscillations as a function of $N_{\sigma}$ relative {to the $\ln N_{\sigma}$ law obeyed by}
the noninteracting case. {However, recall that our results for small even values of~$N_{\sigma}$ are contaminated
with contributions that break translation invariance due to the presence of the boundaries.}

To characterize the {next-to-leading order} behavior, {a finite-size scaling law for fermions}
was {proposed for~$n\ll N_{\sigma}$ in Ref.~\cite{2017arXiv170310587B,*PhysRevA.78.042326}:
\beq
S_{\alpha}(n,N_{\sigma}) = \ln \binom{N_{\sigma}}{n} + \lambda_n^{(\alpha)} + \mathcal{O}(N^{-\delta})\,\label{eq:subcorr}
\eeq
with~$\delta >0$.
Our QMC calculations agree very well with this form, which further confirms the derivations of Ref.~\cite{2017arXiv170310587B}.
Dropping higher-order corrections
in Eq.~\eqref{eq:subcorr}, we estimate the $N$-independent offsets~$\lambda_1^{(\alpha)}$
for~$\alpha=1$ and~$\alpha=2$ by fitting our numerical data for
the respective entropies to the scaling law~\eqref{eq:subcorr}.
The results for the $N$-independent offsets for different values of the coupling can be found in~{Tab.~\ref{Table:Subleading}.}
{Note that the one-body density matrix narrows progressively
such that~$\rho_1(x,x^{\prime})\sim \rho_1(x,x)\delta_{x,x^{\prime}}\sim\rho_1(0,0)\delta_{x,x^{\prime}}$ with increasing coupling.
As can be seen in Tab.~\ref{Table:Subleading},~$\lambda_n^{(\alpha)}$ becomes larger relative to the non-interacting system for increasing coupling
which may be traced back to the dominance of
the diagonal elements~$\rho_1(x,x)$ of the one-body density matrix in this case.
In other words, the entropies increase by interaction effects.
Within} our accuracy, the next-to-next-to-leading order corrections are not resolved, see also Fig.~\ref{Fig:Salpha}.}

The procedure to obtain the uncertainties associated with our results for the entanglement entropy was carried out in
a specifically designed way, as the diagonalization of the one-body density matrix makes error propagation cumbersome and
generally unreliable. The procedure we utilized instead consisted of taking $100$ samples of the one-body density matrix around its QMC average
with the associated statistical uncertainty (assumed to be gaussian), and then calculating the entanglement entropy for every sample so obtained.
Averaging over those samples allowed us to estimate the statistical uncertainties propagated from the one-body density matrix to the
entanglement entropies.

\begin{figure}[t]
\includegraphics[width=\columnwidth]{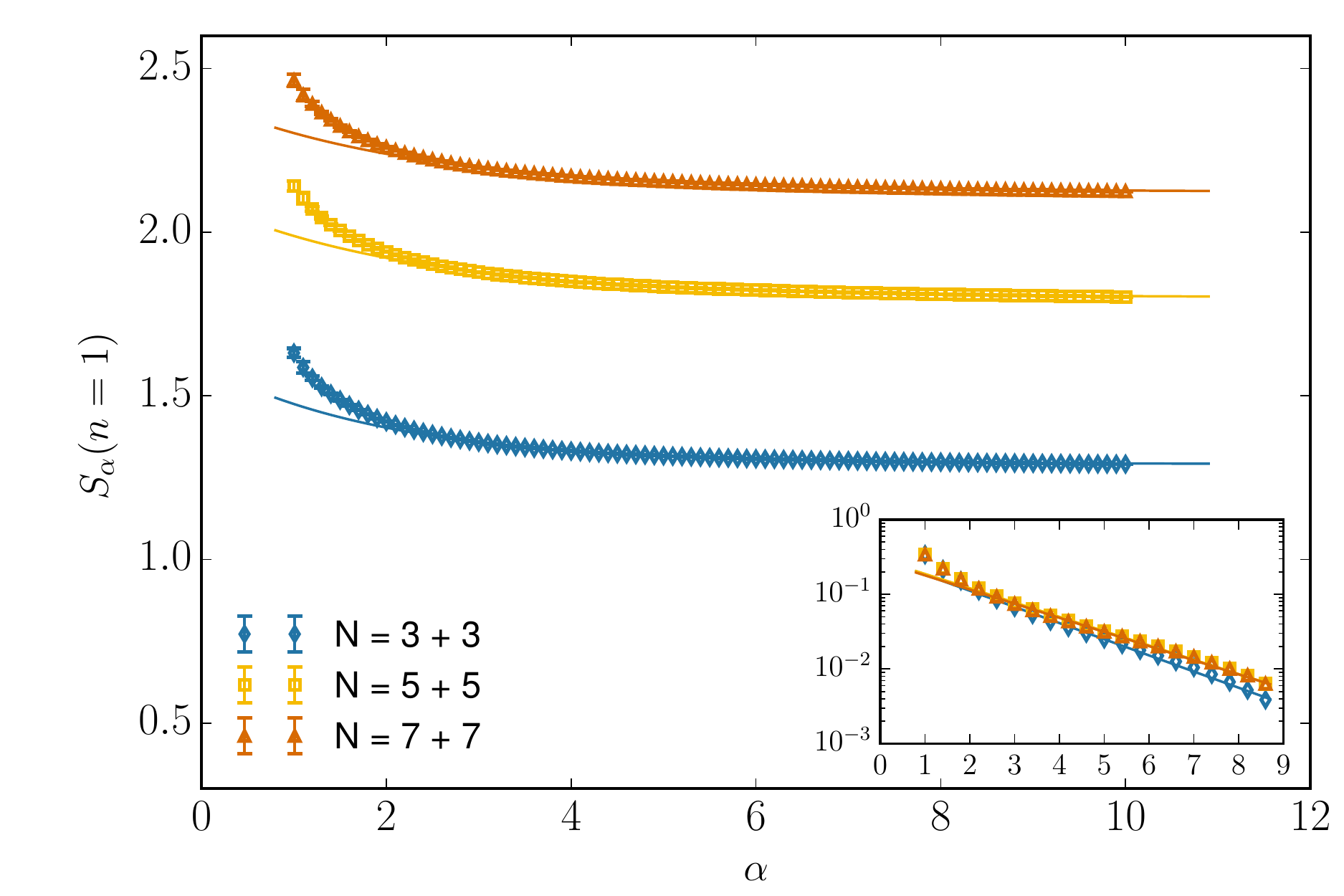}
 \caption{\label{Fig:SalphaVsalpha} Particle-partition entanglement entropy $S_\alpha$ of 1D fermions as a function of $\alpha$ for
three fixed particle numbers $N=3+3,5+5,7+7$ at $\gamma=3.0$ and fixed partition $n=1$. {The {\it von Neumann} result} is shown at $\alpha = 1$. {Solid lines reflect an exponential fit, indicating an
exponential decay of the {\it R\'enyi} entropy for $\alpha \gtrsim 2.0$. To underline this behavior, the same data is shown on a log-scale in the inset.}}
\end{figure}

{In Fig.~\ref{Fig:SalphaVsalpha}, we present our results} for $S_\alpha$ as a function of $\alpha$ for fixed particle numbers
$N=3+3,5+5,7+7$ at $\gamma=2.0$ and fixed partition $n=1$. As in calculations of spatial entanglement in higher dimensions~\cite{UFGEE}, we
find that the large-$\alpha$ limit of $S_\alpha$ is reached rather quickly, as the variation between $\alpha=2$ and $\alpha=5$ is within $5\%$ of
the value at $\alpha=2$ for every case we explored. {In fact,
we observe an exponential decay of the {form
\beq
S_\alpha = S_\infty + S_0\, {\rm e}^{-\frac{\alpha}{\alpha_0}}\,,
\eeq
which} can be appreciated in the inset of Fig.~\ref{Fig:SalphaVsalpha}.}
%

%%%%%%%%%%%%%%%%%%%%%%%%%%%%%%%%%%%%%%%%%%%%%
\section{Conclusions}

{Motivated by experimental and computational advances, we continued here
our lattice QMC benchmarking of {\it Fermi} gases in 1D. Specifically, we
applied lattice QMC methods to the calculation of the one-body density matrix and the associated
single-particle momentum distribution, as well as the pair-correlation function and its associated
momentum distribution. From the single-particle momentum distribution, we extracted
estimates for the speed of sound as well as {\it Tan}'s contact.
We studied systems at a fixed lattice size of $N_x=80$ and presented results
for systems with varying particle content across a wide range of attractive couplings.
We found that systems with small, even particle numbers per species display significant finite-size
effects. This can be traced back to the fact that the ground state in this case
breaks translation invariance in a periodic box, unlike systems with odd particle
numbers per species. The latter appear to converge rapidly to the thermodynamic limit.

In general, our results are in line with earlier studies of two-component gases of fermions in 1D
with an attractive contact interaction between the components. In particular, we have illustrated the
excellent agreement of our results for the ground-state energy with the exact results from the
{\it Bethe} ansatz. Moreover, we have found very good agreement for the contact parameter calculated
differently in a previous study~\cite{GCS1D}. Even more, our first comparatively crude estimates for the
speed of sound are in accordance with the well-known exact results of this
quantity~\cite{Krivnov,PhysRev.130.1605,2004PhRvL..93i0408F}.
Also in accordance with previous studies~\cite{2004PhRvL..93i0408F,wadati},
we find that, in the weakly-coupled regime with~$\gamma \lesssim 1$, the calculated
correlation functions, and therefore also the momentum distributions, are still
well approximated by the ones of the non-interacting system. The dynamics in this regime still appears to
be dominated by the presence of the {\it Fermi} points. In the strongly-coupled regime with~$\gamma\gtrsim 1$,
we then find that the single-particle momentum distributions start to flatten out and the
pair-momentum distribution develops a pronounced maximum at vanishing pair momentum
{relative to the corresponding distribution of the non-interacting system.}
Our estimates for the sound velocity reveal that the system undergoes a crossover from the
weakly-coupled regime, where the sound velocity remains close to the {\it Fermi} velocity,
to a strongly-coupled regime for~$\gamma\gtrsim 2$, where the sound velocity drops drastically.
In detailed analytic studies of the many-body
phase diagram~\cite{PhysRev.130.1605,Krivnov,2004PhRvL..93i0408F,wadati}, this behavior was
traced back to the fact that the dynamics
of the system is governed by {\it Cooper}-type pairing in the weak-coupling limit
and by the formation of tight bosonic molecules in the strong-coupling limit.

Finally, we have used our results for the one-body density matrix to provide testable
predictions for {the {\it R{\'e}nyi}}
{and {\it von Neumann} particle} partition entanglement entropies $S_{\alpha}$
for a partition of $(n=1,n'=N_{\sigma}-1)$ particles.} Our calculations, for varying couplings, orders $\alpha$, and
total particle numbers $N$, {display a logarithmic growth} with $N_{\sigma}$ and mild
oscillations on top of that growth, which further confirms the results of recent analytic studies.
Additionally, we explored the $\alpha$-dependence of $S_{\alpha}(N)$ for several $N$ and found that it decays
exponentially to the limiting value $S_\infty(N)$ {with an approximately $N$-independent} decay amplitude and rate.
%The specific value of $S_\infty(N)$ is governed completely by the largest eigenvalue of the one-body density matrix
%and the exponential decay can be captured by considering just two different eigenvalues.
%However, an attempt to use those two eigenvalues to explain the observed decay fails, which leads
%us to conclude that multiple eigenvalues of similar magnitude contribute to the overall behavior as a function
%{of $\alpha$, which is also consistent with our observation that the one-body matrix narrows progressively
%with increasing coupling strength.}

{One of the goals of the present work is to benchmark our lattice QMC approach for the
computation of ground-state properties with known exact results. Indeed, we have
found very good agreement with the exact results for the observables considered here.
Our present study therefore sets the methodological stage for future studies of correlation functions
of {\it Fermi} gases in higher dimensions, where exact results for, e.g., correlation functions,
are urgently needed. In particular, for studies of systems with a finite spin- and mass-imbalance,
the computation of general correlation functions is of interest as the formation of an
{inhomogeneous (LOFF-type) ground state} is expected to leave its imprint on
these quantities.}
%

%%%%%%%%%%%%%%%%%%%%%%%%%%%%%%%%%%%%%%%%%%%%%
{
{\it Acknowledgments.--} J.B. and L.R. are grateful to H.-W.~Hammer
and A.~Volosniev for useful discussions {and are grateful to A.~Volosniev for comments on the manuscript}.
This work was supported by HIC for FAIR within the LOEWE program of the State of Hesse.
Moreover, this material was based upon work supported by the
National Science Foundation under Grant No.
PHY{1452635} (Computational Physics Program).
}

%%%%%%%%%%%%%%%%%%%%%%%%%%%%%%%%%%%%%%%%%%%%%
\appendix
%%%%%%%%%%%%%%%%%%%%%%%%%%%%%%%%%%%%%%%%%%%%%
\section{Systematics \label{app:a}}
{In this appendix} we discuss the behavior
of our results as a function of the lattice size $N_x$ and the imaginary projection time $\beta$.
Although, in principle, one should compute quantities in the limit of infinite $N_x$ and $\beta$ to
guarantee full convergence to the respective physical limit, it is often sufficient to carry out calculations at
finite, but large, parameter values, as shown below. Comparisons between different systems are shown for a representative
system {with $N=8+8$ particles} with strong and weak interactions.
{As mentioned in the main text, the shown error bars reflect statistical uncertainties of $\sim 5000$ samples.
The associated typical autocorrelation times for the energy are of the order of $10^{-2}$.
In our discussion below,
decorrelation between points at different $k / \kf$ in the momentum distribution is tacitly assumed. However, an
explicit detailed
analysis of the autocorrelation times for this quantity has not been performed
and therefore the true error bar on the results may be (slightly) underestimated as the autocorrelation
time may have been underestimated for this quantity.}

\subsection{Finite lattice size}
\begin{figure*}[t]
 \includegraphics[width=\columnwidth]{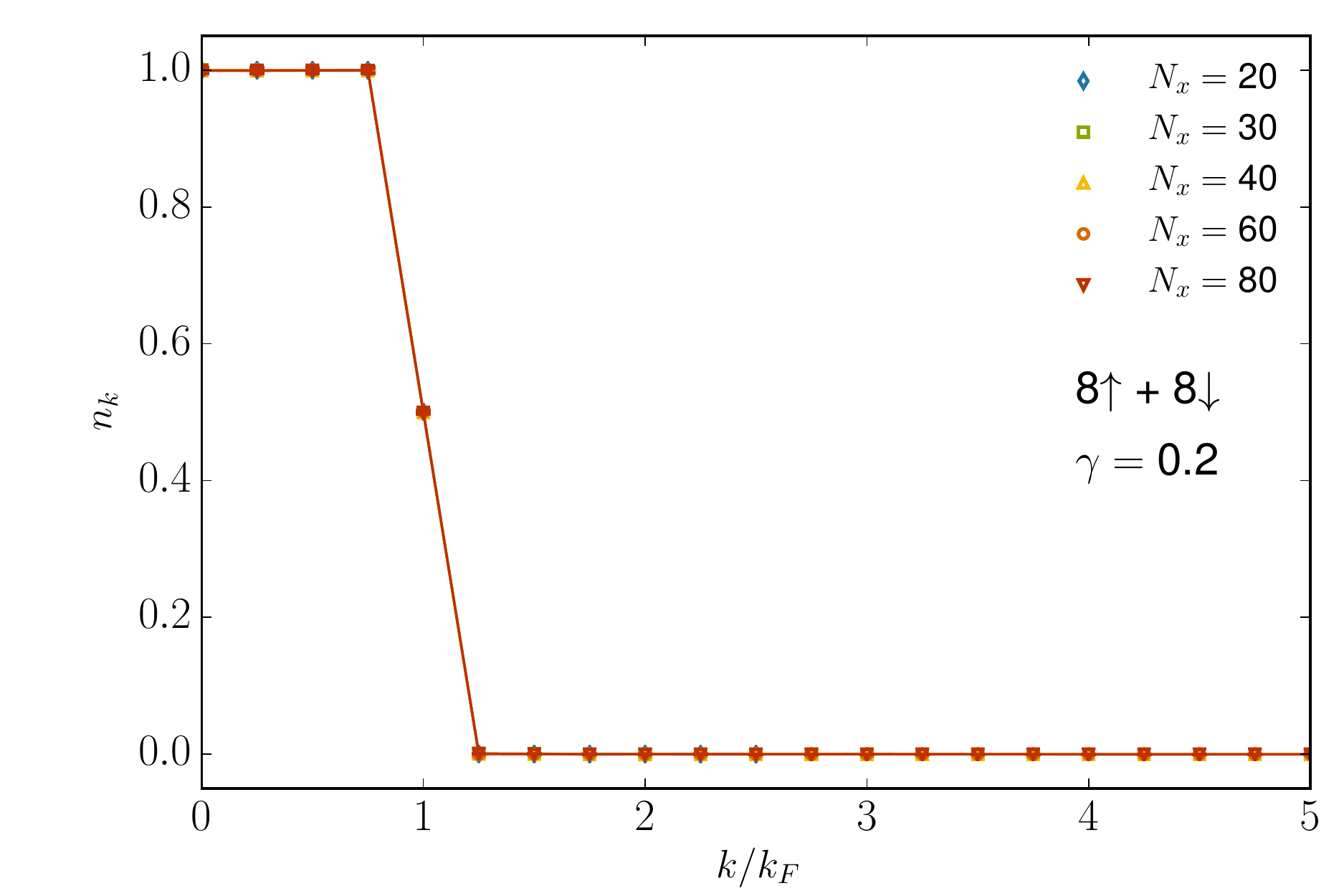}
 \includegraphics[width=\columnwidth]{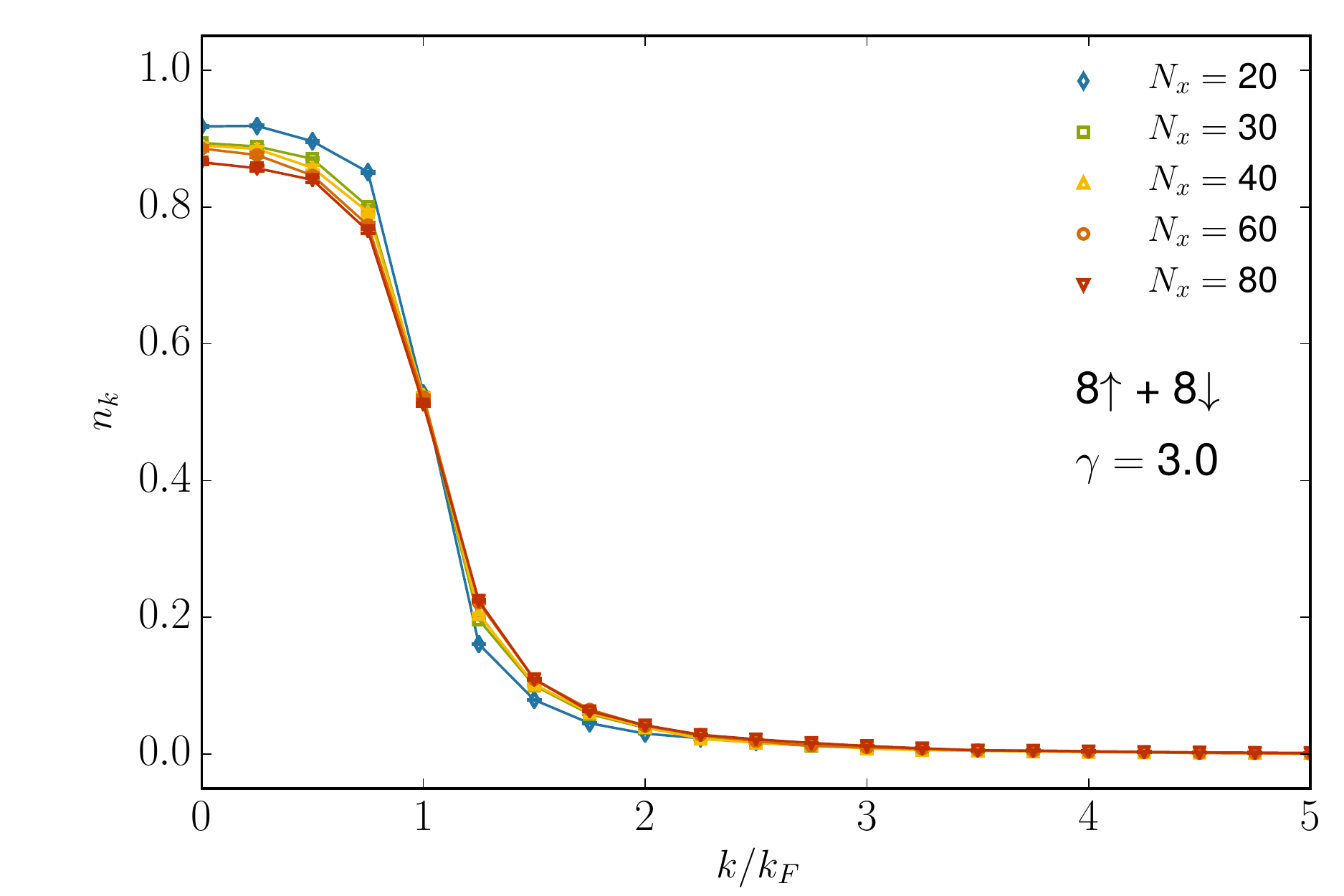}
 \hspace*{0.3cm}\includegraphics[width=2.0\columnwidth]{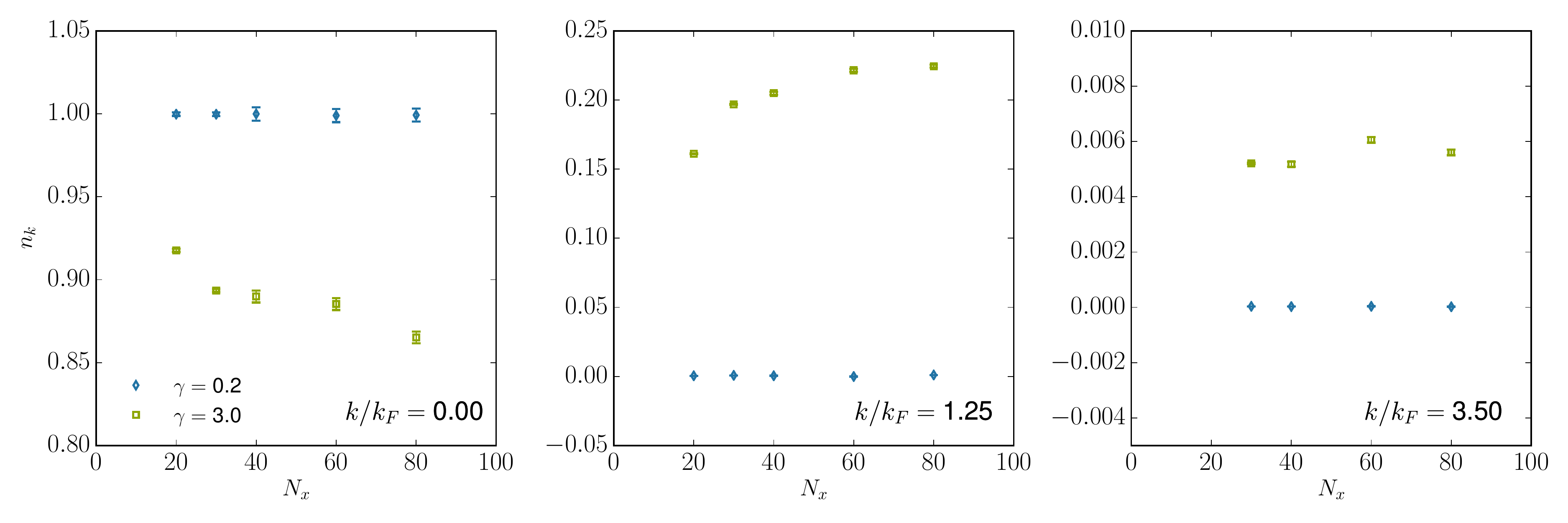}
 \caption{\label{Fig:finiteLattice} Top panels: Momentum distribution~$n_k$ as a function of~$k/\kf$
 for values of the lattice size $N_x = 20, 30, 40, 60$, and $80$.
 {The left panel shows the very weakly coupled case of $\gamma = 0.2$. Strongly coupled systems at $\gamma = 3.0$ are shown in the right
 panel. In the latter case, finite-volume effects are clearly more pronounced, albeit still comparatively small. Bottom panels: Momentum distribution~$n_k$
 as a function of the lattice size~$N_x$ for~$k/\kf=0, 1.25, 3.5$ (from left to right) for $\gamma=0.2$ and~$\gamma=3.0$.}}
\end{figure*}

{In the present work, we employ periodic boundary conditions in our calculations.
We shall now address the effect of finite lattice sizes in case of such boundary conditions in} order to have a more complete
overview of possible systematic errors.
To this end, we exploit the fact that, in the infinite-volume limit,
physical observables\footnote{More precisely, observables rendered dimensionless
with suitably chosen powers of the density (or, equivalently, the {\it Fermi} momentum).} should only depend on the value of the dimensionless coupling~$\gamma=g/n$.
This implies that our results for a given
fixed value of the dimensionless coupling~$\gamma$ should not exhibit an explicit dependence on the density~$n=N/L$, if the box size~$L$ has been
chosen sufficiently large.
In  the top panels of Fig.~\ref{Fig:finiteLattice}, we show the momentum distribution $n_k$ for lattice sizes of $N_x = 20, 30, 40, 60$, and $80$
for fixed~$\gamma=0.2$ and~$\gamma=3.0$. Recall that~$N_x=L/\ell$ where~$\ell$ is the lattice spacing.
In the bottom panels of Fig.~\ref{Fig:finiteLattice}, we illustrate the scaling behavior of~$n_k$ with the lattice size~$N_x$ for $k/\kf=0, 1.2, 3.5$ (from left to right), again
for fixed~$\gamma=0.2$ and~$\gamma=3.0$.
We clearly observe that weakly coupled systems,
exemplified by the momentum distribution with $\gamma=0.2$, show a indiscernible
dependence on the lattice size and therefore associated errors are almost absent, as evident
from Fig.~\ref{Fig:finiteLattice}. For the strongly coupled case at $\gamma=3.0$, finite-size effects are more
pronounced, most prominently for momenta~$k$ close to the {\it Fermi} point~$\kf$. However, even close to the {\it Fermi} point, the volume dependence
is already very weak for~$N_x\gtrsim 60$. Given these results, we do not employ an extrapolation to the infinite-volume limit but consider {$N_x = 80$} to be sufficiently converged, which
is therefore the value used throughout this work.

Obviously, the momentum distribution is not the only quantity influenced by finite-size effects.
{Our results for the one- and two-body density matrices, however, feature} the
same type of convergence and will not be considered separately at this point. We add that we also find agreement with previous
studies of the ground-state energy in 1D~\cite{GCS1D}.

%%%%%%%%%%%%%%%
%%%%%%%%%%%%%%%

%
\subsection{Finite imaginary time}
%

%%%%%%%%%%%%%%%
\begin{figure*}[t]
 \includegraphics[width=\columnwidth]{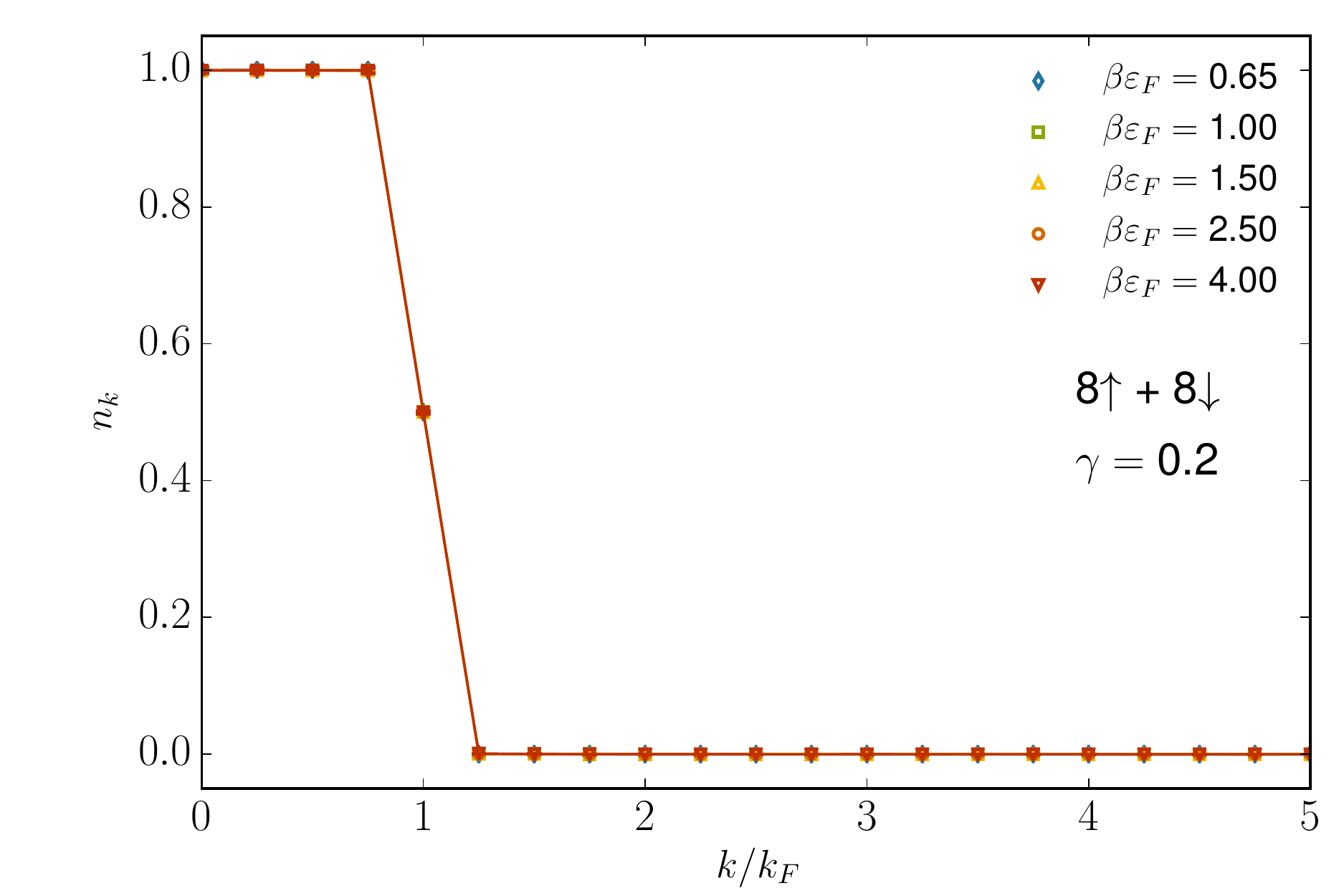}
 \includegraphics[width=\columnwidth]{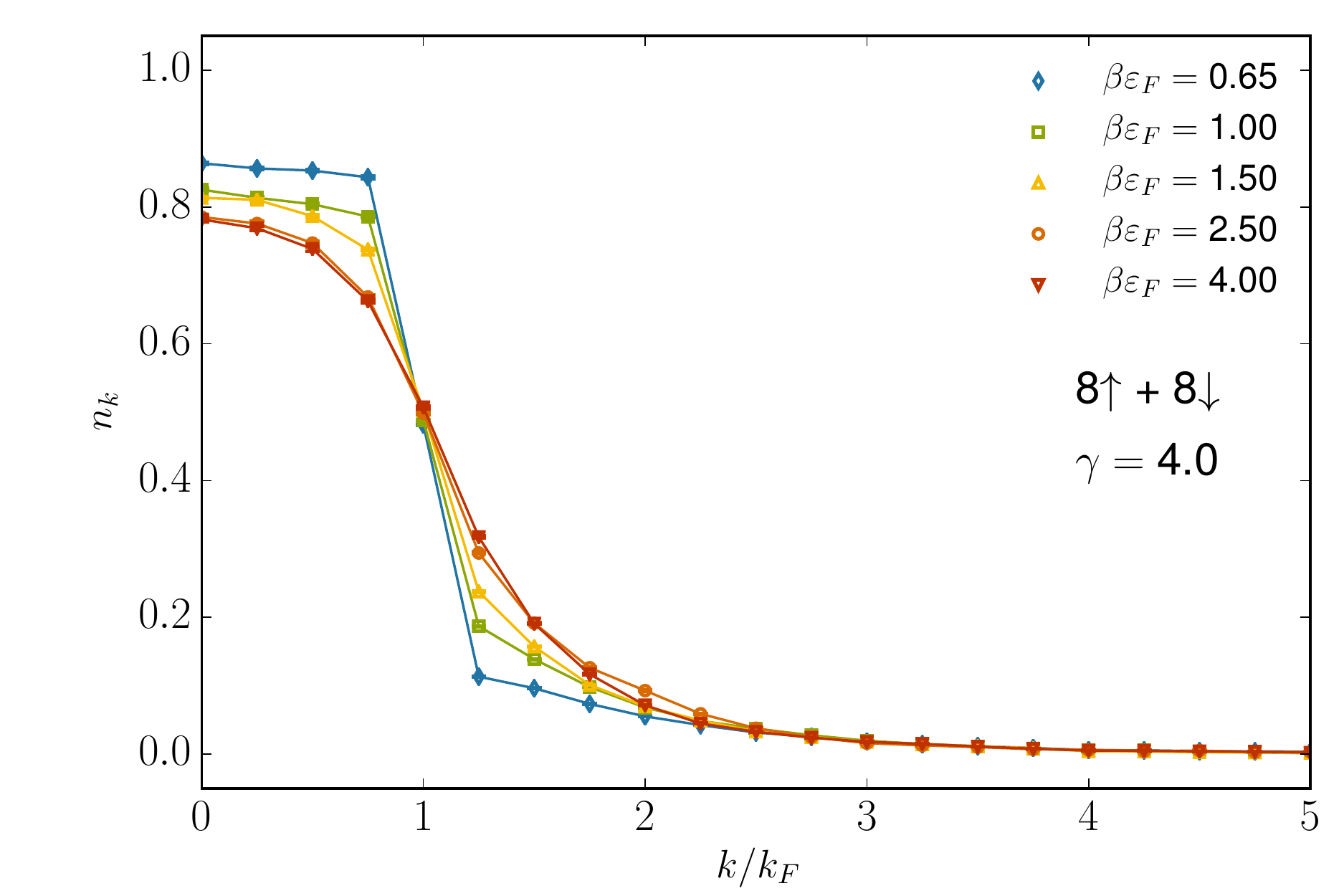}
 	\hspace*{0.3cm}\includegraphics[width=2.0\columnwidth]{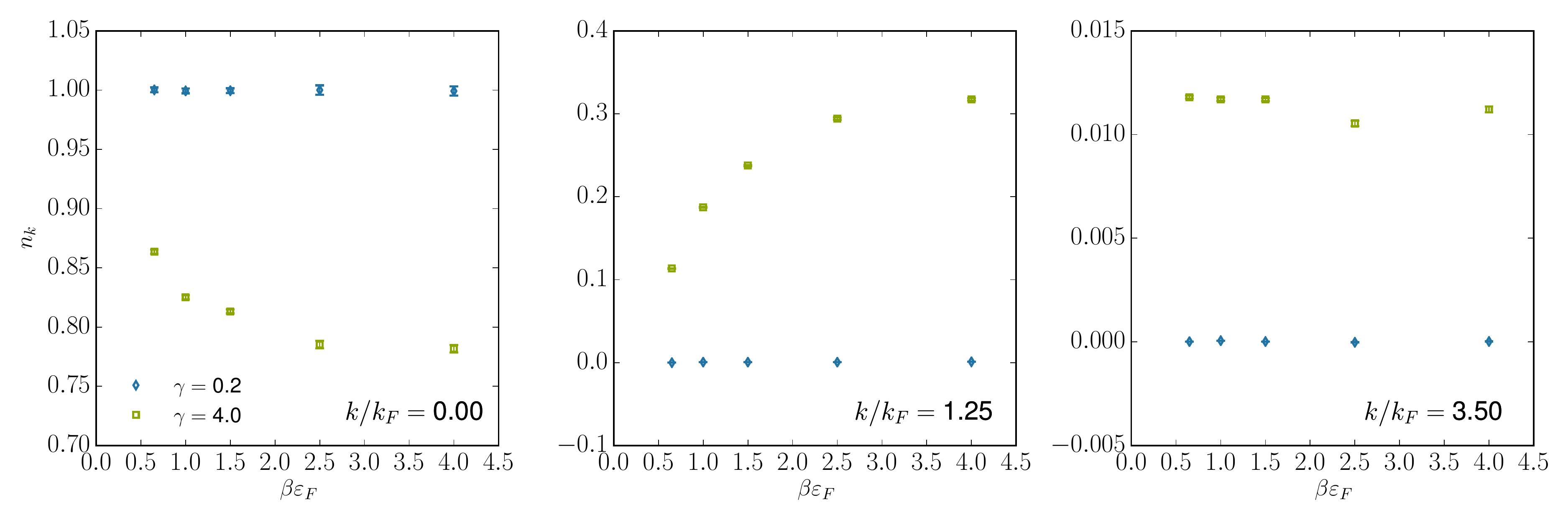}
 \caption{\label{Fig:finiteBeta} Top panels: Momentum distribution~$n_k$ as a function of~$k/\kf$
 for different values of the extent of the imaginary time axis~$\beta\epsilon_{\rm F} = 0.65, 1.0, 1.5, 2.5$ and $4.0$.
 The weakly interacting system with $\gamma = 0.2$ {(top left panel) is
 immediately converged. In the strongly coupled case  at $\gamma = 4.0$ (top right panel), the results converge slower as
 a function of $\beta \epsilon_{\rm F}$.
 Bottom panels: Momentum distribution~$n_k$
 as a function of~$\beta\epsilon_{\rm F}$ for~$k/\kf=0, 1.25, 3.5$ (from left to right) for $\gamma=0.2$ and~$\gamma=3.0$.}}
\end{figure*}
%%%%%%%%%%%%%%%

As mentioned in the text, we evaluate the projection up to a finite value of $\beta$,
corresponding to a finite effective inverse temperature. Since our approach
exploits an initial guess state (in our case taken to be a {\it Slater} determinant) and
projects to the ground-state, we need to make sure that the obtained results are fully
converged to the limit $\beta\to\infty$. {In Fig.~\ref{Fig:finiteBeta}, we show this effect for
two systems in the weakly and {strongly interacting regime. Again, as expected, the essentially
free case at $\gamma = 0.2$ shows no dependence on $\beta$ and is converged almost immediately, see also the bottom panels
of Fig.~\ref{Fig:finiteBeta} where~$n_k$ is shown as a function of~$\beta\epsilon_{\rm F}$.
For strongly interacting systems, larger projection times are required to observe convergence. As
depicted in the bottom panels of
Fig.~\ref{Fig:finiteBeta}, we observe that the dependence on $\beta\epsilon_{\rm F}$ starts to become
weak for~$\beta \epsilon_{\rm F} \simeq 2.5$, even for
the most strongly coupled systems considered in this work. Therefore, we have not employed an extrapolation to infinite~$\beta\epsilon_{\rm F}$
but have rather used~$\beta \epsilon_{\rm F} \simeq 2.5$ to obtain the results presented in the main part of this work.}}

%%%%%%%%%%%%%%%%%%%%%%%%%%%%%%%
\bibliography{refs_1dzt}

\end{document}